\documentclass[%
 reprint,
superscriptaddress,
nofootinbib,
amsmath,amssymb,
aps,
]{revtex4-2}

\usepackage{graphicx}            
\usepackage{pgfplots}
\pgfplotsset{compat=1.18}
\usetikzlibrary{plotmarks}

\definecolor{mplblue}  {RGB}{  0,  0,255}
\definecolor{mplgreen} {RGB}{  0,128,  0}
\definecolor{mplred}   {RGB}{255,  0,  0}
\definecolor{mplpurple}{RGB}{128,  0,128}
\definecolor{mplorange}{RGB}{255,165,  0}
\definecolor{mplgray}  {RGB}{128,128,128}
\definecolor{mplbrown} {RGB}{165, 42, 42}   
\definecolor{mplgrid}  {RGB}{176,176,176}   

\newcommand{\curvelw}{3.5pt}    
\newcommand{\mwpmlw} {1.75pt}   
\newcommand{\errlw}  {2.5pt}    
\newcommand{\msize}  {4.5pt}    
\newcommand{\mwpmms} {2.25pt}   
\newcommand{\mewidth}{1.2pt}    

\tikzset{
  mpldotted/.style  ={dash pattern=on 3.5pt off 5.8pt},
  mpldashdot/.style ={dash pattern=on 22.4pt off 5.6pt on 3.5pt off 5.6pt},
}

\pgfplotsset{
  dcurve/.style={
    color=#1,
    line width=\curvelw,
    mark=*,
    mark size=\msize,
    mark options={fill=#1, draw=#1, line width=\mewidth},
    error bars/y dir=both,
    error bars/y explicit,
    error bars/error bar style={color=#1, line width=\errlw},
    error bars/error mark=none,
  },
  dkey/.style={
    color=#1, line width=\curvelw, mark=*, mark size=\msize,
    mark options={fill=#1, draw=#1, line width=\mewidth},
  },
}
\usepackage{graphicx}
\usepackage{dcolumn}
\usepackage{bm}
\usepackage{algorithm2e}
\usepackage{tikz}
\usepackage{flushend}
\usepackage{todonotes}
\usepackage[normalem]{ulem}

\definecolor{luca}{rgb}{0.8, 0.1, 0.1}

\usepackage{hyperref}
\usepackage{amsthm}
\usepackage{comment}
\usepackage{enumitem}
\usepackage{mathtools}
\usepackage{natbib}
\usepackage{amsfonts}
\usepackage{amssymb}
\usepackage[T1]{fontenc}

\newcommand{\mf}{\mathcal F}
\newcommand{\mg}{\mathcal G}

\newcommand{\me}{\mathcal E}
\newcommand{\med}{{\mathcal E}_d}
\newcommand{\mv}{\mathcal V}
\newcommand{\mvd}{{\mathcal V}_d}

\newcommand{\smin}{\setminus}
\newcommand{\nn}{\mathcal{N}}

\newcommand{\graph}{\mathcal{F}}
\newcommand{\setvnodes}{\mathcal{V}}
\newcommand{\setcnodes}{\mathcal{C}}
\newcommand{\setedges}{\mathcal{A}}

\newcommand{\vnode}{\mathtt{v}}
\newcommand{\cnode}{\mathtt{c}}

\newcommand{\prior}{\rho}
\newcommand{\priorlog}{\ell}
\newcommand{\posteriorlog}{\lambda}

\newcommand{\mvc}{m_{\vnode \to \cnode}}
\newcommand{\mcv}{m_{\cnode \to \vnode}}

\newcommand{\quot}[1]{%
  #1/{\sim}%
  }

\newcommand{\weight}{\omega}

\usepackage{bm}
\usepackage {tikz}
\usetikzlibrary {positioning}
\usetikzlibrary{arrows.meta}
\usetikzlibrary{decorations.pathreplacing}

\tikzset{small dot/.style={fill=black,circle,scale=1.5}}
\tikzset{small dot 1/.style={fill=red!40,circle,scale=1.5}}
\tikzset{small dot 2/.style={fill=blue!40,circle,scale=1.5}}
\tikzset{small dot 3/.style={fill=green!40,circle,scale=1.5}}
\tikzset{small dot 4/.style={fill=red!70,circle,scale=1.5}}
\tikzset{small dot 5/.style={fill=blue!70,circle,scale=1.5}}
\tikzset{small dot 6/.style={fill=green!70,circle,scale=1.5}}

\usepackage{caption}
\usepackage{subcaption}
\usepackage{bbm}

\usepackage{lmodern}
\usepackage{physics}

\usetikzlibrary{fit}

\usepackage{graphicx}
\usepackage{subcaption}

\begin{document}

\title{Achieving Thresholds via Standalone Belief Propagation on Surface Codes}

\author{Pedro Hack}
\email{pedro.hack@dlr.de}
\affiliation{German Aerospace Center, Germany}

\author{Luca Menti}
\email{luca.menti@dlr.de}
\affiliation{German Aerospace Center, Germany}

\author{Francisco Lázaro}
\email{francisco.lazaroblasco@dlr.de}
\affiliation{German Aerospace Center, Germany}

\author{Alexandru Paler}
\email{alexandru.paler@aalto.fi}
\affiliation{Aalto University, Finland}

\begin{abstract}
The standard approach to Belief Propagation (BP) decoding passes messages on the Tanner graph of the quantum error-correcting (QEC) code, and is known to lack a threshold for the surface code. We propose novel BP decoders that exchange messages on the decoding graph and obtain thresholds (both for code capacity and circuit-level noise) via standalone BP for the surface code under depolarizing noise. Our approach, similarly to the minimum weight perfect matching (MWPM) decoder, is applicable to any graphlike QEC code. The thresholds observed with our decoders are close to those obtained by MWPM. This result opens the path towards scalable hardware-accelerated implementations of MWPM-compatible decoders.
\end{abstract}

\providecommand{\keywords}[1]
{
  \small	
  \textbf{\textit{Keywords---}} #1
}

\keywords{one, two, three, four}

\maketitle

\section{Introduction}

Standard belief propagation (BP) is known to lack a threshold for the surface code, a failure usually attributed to the highly loopy structure of the code's Tanner graph.
In this work we show that this failure is not intrinsic to BP, but a consequence of the graph on which messages are passed. We move the message-passing from the code's Tanner graph to the decoding graph, the graph on which MWPM operates, and recover threshold behavior for the surface code, despite the presence of short cycles in the latter graph.

\subsection{Motivation}

For graphlike CSS codes, where each physical qubit participates in at most two X- and two Z-checks, the standard decoding algorithm is minimum-weight perfect matching (MWPM).
Graphlike CSS codes include several important codes~\cite{dennis2002topological,breuckmann2016constructions,bravyi2012subsystem,suchara2011constructions,bacon2006operator,hastings2021dynamically} such as the surface code. At the same time, graphlike approaches like MWPM can be adapted to more general codes~\cite{bombin2006topological,kubica2023efficient,brown2023conservation}.  

The complexity of a naïve implementation of MWPM scales like $O(n^3 \log n)$, where $n$ is the number of physical qubits \cite{higgott2022pymatching}. In general, this is considered to be prohibitively slow for decoding large scale fault-tolerant quantum computations, which hence motivates the study of fast MWPM implementations whose performance does not degrade abruptly~\cite{fowler2013minimum,delfosse2021almost,higgott2025sparse}.

Herein, we explore the use of standalone BP-based approaches to MWPM as a decoder for quantum error correction. We present algorithms which save computational time when compared to MWPM by taking a BP approach to the problem, that is, by \textbf{running message-passing on the QEC decoding graph}, the same graph where MWPM operates. This contrasts with the standard BP approach to decoding~\cite{kuo2020refined,liu2019neural}, which performs message-passing directly on the QEC code's Tanner graph.

Compared to the standard approach of running BP on the Tanner graph, our algorithms have slightly higher time complexity owing to the larger size of the graph on which they operate (Sec.~\ref{sec:methods}).
However, they can significantly improve the performance of BP decoding, and in contrast to vanilla BP~\cite{demarti2024decoding} which does not achieve threshold, our algorithms are achieving it (Sec.~\ref{sec:res}).

While message-passing algorithms for computing matchings have been studied extensively~\cite{bayati2005maximum,sanghavi2007linear,bayati2008max,gelfand2013belief,ahn2015minimum}, their application to quantum error correction has not been explored, and empirical evidence of their practical performance remains limited. Moreover, prior work primarily focused on approximating MWPM solutions, whereas our focus is on evaluating their performance under constrained time complexity.

\subsection{MWPM on the surface code}

Our BP-based method is applicable as a direct replacement of vanilla MWPM. For simplicity, we focus on uncorrelated decoding in the surface code, and decode $X$ and $Z$ Pauli errors separately. In the following, we fix the case of $Z$ Pauli errors and refer to the $X$ parity check matrix by $\bm{H}$. Moreover, we explain our method using the code capacity noise model. This means that what we call check refers to a detection event.

MWPM works on the \textbf{decoding graph} $\mg=(\mvd,\med)$. For each error realization $\bm{e}$, the decoding graph consists of one node $i \in \mvd$ for each unsatisfied check, as well as one edge $(i,j) \in \med$ for each pair of unsatisfied checks. The edge $(i,j)$ represents the fact that one can map the unsatisfied check associated to $i$ to the one associated to $j$. Additionally, for each unsatisfied check associated with some vertex $i \in \mvd$, the decoding graph contains a \textbf{boundary node} $i_b \in \mvd$ and a boundary edge $(i,i_b) \in \med$. The boundary node and the edge represent the fact that one can map the unsatisfied check to the boundary.

For a given error realization $\bm{e}$, we will denote by $s$ its associated number of unsatisfied checks. In this scenario, the decoding graphs consists of $|\mvd| = 2s$ nodes and $|\med| = s(s-1)/2+ s$ edges. For a given physical error rate $p$, and for large $n$, we expect to have $pn$ faulty qubits. If that is the case, then the maximum expected number of unsatisfied checks is $2pn=O(n)$. This takes place whenever the faulty qubits do not share any check with each other, so each of them triggers two checks in a graph-like code.

\begin{figure*}[!t]
\centering

\begin{subfigure}[b]{0.2\textwidth}
  \centering
  \includegraphics[width=\columnwidth]{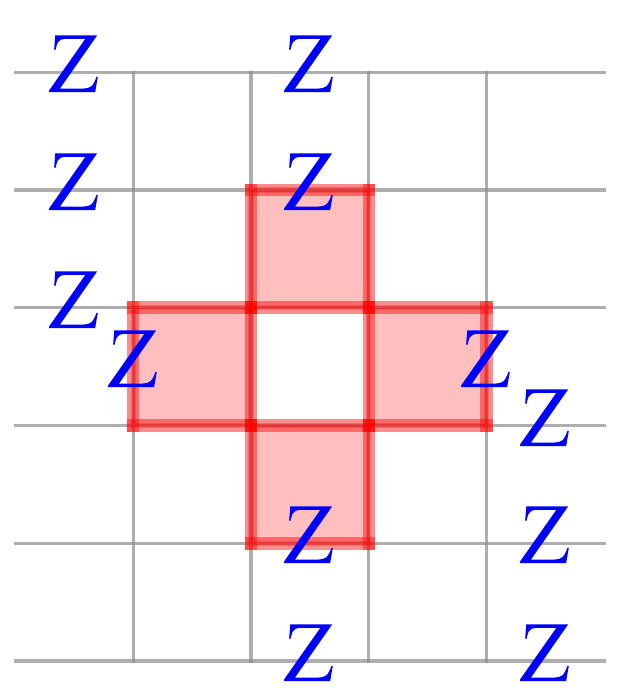}
  \caption{}
  \label{fig: surface code syndrome}
\end{subfigure}
\hfil
\begin{subfigure}[b]{0.26\textwidth}
  \centering
  \resizebox{\columnwidth}{!}{
\begin{tikzpicture}[
  x=1.45cm, y=0.95cm,          
  rednode/.style={
    circle, fill=red!80!black, draw=red!80!black,
    minimum size=9pt, inner sep=0pt
  },
  orangenode/.style={
    circle, fill=orange!90!yellow, draw=orange!70!red,
    minimum size=9pt, inner sep=0pt
  },
  edge/.style={thick, black},
  font=\large
]

\coordinate (S2) at ( 0.00,  2.20);
\coordinate (S1) at (-1.30,  0.00);
\coordinate (S3) at ( 1.30,  0.00);
\coordinate (S4) at ( 0.00, -2.20);

\coordinate (L1) at (-2.80,  3.60);
\coordinate (L2) at ( 0.00,  3.60);
\coordinate (L3) at ( 2.90, -3.60);
\coordinate (L4) at ( 0.00, -3.60);

\draw[edge] (S1) -- (S2);
\draw[edge] (S1) -- (S3);
\draw[edge] (S1) -- (S4);
\draw[edge] (S2) -- (S3);
\draw[edge] (S2) -- (S4);
\draw[edge] (S3) -- (S4);

\draw[edge] (S1) -- (L1);
\draw[edge] (S2) -- (L2);
\draw[edge] (S3) -- (L3);
\draw[edge] (S4) -- (L4);

\foreach \n in {S1,S2,S3,S4} { \node[rednode]    at (\n) {}; }
\foreach \n in {L1,L2,L3,L4} { \node[orangenode] at (\n) {}; }

\node[red] at (-1.80,  0.00) {$s_1$};
\node[red] at (-0.30,  2.55) {$s_2$};
\node[red] at ( 1.75,  0.00) {$s_3$};
\node[red] at ( 0.30, -2.45) {$s_4$};

\node[blue] at (-2.35,  1.50) {$w_{1,1}$};
\node[blue] at ( 0.40,  3.00) {$w_{2,2}$};
\node[blue] at ( 2.70, -2.30) {$w_{3,3}$};
\node[blue] at ( 0.40, -3.15) {$w_{4,4}$};

\node[blue] at (-1.05,  1.35) {$w_{1,2}$};
\node[blue] at (-0.55, -0.28) {$w_{1,3}$};
\node[blue] at (-0.95, -1.35) {$w_{1,4}$};
\node[blue] at ( 1.05,  1.35) {$w_{2,3}$};
\node[blue] at ( 0.65,  0.20) {$w_{2,4}$};
\node[blue] at ( 1.05, -1.40) {$w_{3,4}$};

\end{tikzpicture}
}
  \caption{}
  \label{fig: decoding graph}
\end{subfigure}


\begin{subfigure}[b]{0.75\textwidth}
  \centering
  \scalebox{0.5}{
\begin{tikzpicture}[
    x=1.3cm, y=5cm,
    factor/.style={rectangle, draw=black!80, thick, minimum size=8mm, font=\normalsize},
    var/.style={circle, draw=black!80, thick, minimum size=5mm, font=\normalsize},
    edge/.style={thick, draw=black!70},
    prior/.style={->, >=stealth, thick, draw=black!70}
]
    \normalsize

    \def\dvar{1.35}   
    \def\dfac{1.90}   
    \def\xc{6.5}      

    \foreach \nm/\i/\lab in {%
        11/1/{1,1}, 12/2/{1,2}, 13/3/{1,3}, 14/4/{1,4},
        22/5/{2,2}, 23/6/{2,3}, 24/7/{2,4},
        33/8/{3,3}, 34/9/{3,4},
        44/10/{4,4}%
    }{
        \pgfmathsetmacro{\xv}{\xc + (\i-5.5)*\dvar}
        \node[var] (v\nm) at (\xv, 0) {$\vnode_{\lab}$};
        \node[below=0.7cm of v\nm] (l\nm) {$\prior_{\lab}$};
        \draw[prior] (l\nm) -- (v\nm);
    }

    \foreach \j in {1,2,3,4}{
        \pgfmathsetmacro{\xf}{\xc + (\j-2.5)*\dfac}
        \node[factor] (c\j) at (\xf, 1) {$\cnode_\j$};
    }

    \foreach \v in {11, 12, 13, 14} { \draw[edge] (c1) -- (v\v); }
    \foreach \v in {12, 22, 23, 24} { \draw[edge] (c2) -- (v\v); }
    \foreach \v in {13, 23, 33, 34} { \draw[edge] (c3) -- (v\v); }
    \foreach \v in {14, 24, 34, 44} { \draw[edge] (c4) -- (v\v); }

\end{tikzpicture}
}
  \caption{}
  \label{fig: factor graph}
\end{subfigure}

\caption{Decoding and factor graphs. (a) Example error realization $\bm{e}$ in the $d=5$ unrotated surface code. $Z$ phase-flip errors are \textcolor{blue}{blue} and unsatisfied syndrome plaquettes are \textcolor{red}{red}. (b) Decoding graph associated to $\bm{e}$. The unsatisfied syndromes are associated to red nodes and the boundary nodes are orange. We include one edge for each pair of unsatisfied syndromes and one edge that joins each unsatisfied syndrome to the boundary. 
    Edge $(i,j)$ has a weight $\weight_{i,j}\equiv (p/(1-p))^{w_{i,j}}$, where $p$ is the physical error rate and $w_{i,j}$, $i\neq j$, is the weight of the shortest error path that satisfies the syndromes connected by the edge, whereas $w_{i,i}$ is the shortest path from $s_i$ to the boundary.
    (c)~Factor graph associated to $\bm{e}$. Each unsatisfied syndrome $i$ is associated to a factor $c_i$ \eqref{eq: syndrome factors}
    and a variable $\vnode_{i,i}$. Each pair of unsatisfied syndromes $i,j$ ($i<j$)
    is associated to a variable $\vnode_{i,j}$. Each variable $\vnode_{i,j}$, $i \leq j$, is associated to a prior probability $\prior_{ij}$ \eqref{eq: two-syn var factors}. }
\label{fig: different factor graphs}
\end{figure*}

As a result, taking into account the boundary vertices, the decoding graph has at most $|\mvd| = 2pn+2pn=O(n)$ nodes and $|\me| = 2pn(2pn-1)/2+ 2pn = O(n^2)$ edges. Fig.~\ref{fig: different factor graphs} shows an error realization and its associated decoding graph.

MWPM associates a weight to each edge in the decoding graph and runs the Blossom algorithm~\cite{edmonds1965paths,higgott2022pymatching} in order to find an error $\bm{e'}$ with minimal weight that has the observed syndrome $\bm{H} \bm{e'} = \bm{H} \bm{e}$. Instead of doing this, we associate a \textbf{factor graph} $\mf$ to the decoding graph and correct errors by running BP on $\mf$.  

\section{Methods}
\label{sec:methods}

\subsection{Bipartite Graph Representation}



We define the bipartite graph $\graph$ associated to the decoding graph $\mg$ as 
$\graph = (\setvnodes , \setcnodes, \setedges )$ where
\begin{itemize}
    \item $\setvnodes= \{ \vnode_{1,1}, \dots, \vnode_{1, s}, \dots, \vnode_{s,1}, \dots, \vnode_{s, s} \}$ is the set of variable nodes. 
    Variable nodes in the bipartite graph are associated to edges in the decoding graph. In particular, variable node $\vnode_{i,j}$, $i\neq j$, is associated to edge $(i,j)$ in the decoding graph $\mg$, i.e., to an edge connecting the $i$-th and $j$-th unsatisfied syndromes. Similarly,  variable node $\vnode_{i,i}$ is associated to the edge $(i, i_b)$ in $\mg$, i.e., to an edge connecting the $i$-th unsatisfied syndrome to the boundary. Although we have presented them independently, note that $\vnode_{i,j}=\vnode_{j,i}$, i.e. there are $s(s-1)/2+s$ \textbf{independent} variables in $\setvnodes$.
    
    \item $\setcnodes = \{ \cnode_1, \dots, \cnode_s\}$ is the set of factor nodes associated to the unsatisfied syndromes. 
    
    \item $\setedges$ is the set of (undirected) edges of the bipartite graph. The bipartite graph contains $s(s-1)/2+s$ edges, one for each independent variable. In particular, we have that every factor node $\cnode_i$ is connected to the variable nodes $\vnode_{i,1}, \dots, \vnode_{i, s}$, i.e., it has exactly $s$ incident edges. That is, variable node $\vnode_{i,j}$, $i \neq j$, is connected to factor nodes $\cnode_i$ and $\cnode_j$, whereas variable node $\vnode_{i,i}$ is connected only to $\cnode_i$.
\end{itemize}

 The random variables $v_{i,j}$ associated to the variable nodes $\vnode_{i,j}$ take binary values. In particular, we associate:
 \begin{itemize}
     \item the value $v_{i,j}= 1$ to the corresponding edge in the decoding graph $\mg$ being included in the matching;
     \item and the value $v_{i,j}= 0$ to the corresponding edge in $\mg$ not being included in the matching.
 \end{itemize}
For the sake of notational simplicity, we will use $v_{i,j}$ to refer to the binary random variables, as well as their corresponding realizations. 


Each factor node $\cnode_i \in \setcnodes$ in our bipartite graph $\graph$ encodes a local   compatibility constraint. In particular, the constraint is that only one out of the $s$ variable nodes it is connected to can take value one, that is, that each unsatisfied syndrome is invovled in a single matching. Formally:
\begin{equation}
\label{eq: syndrome factors}
    \cnode_i(v_{i,1},\dots,v_{i,s} ) = 
    \begin{cases}
    1 & \mkern-10mu \text{ if } \sum_{j=1}^{s} v_{i,j}=1,\\
    0 & \mkern-10mu \text{ else,}
    \end{cases} \mkern-38mu
\end{equation}

The prior probability associated to random variable $v_{i,j}$, $\prior_{i,j}$, corresponds to 
\begin{equation}
\label{eq: two-syn var factors}
    \prior_{ij} = 
    \begin{cases}
    \left(\frac{
    p}{1-p}\right)^{w_{i,j}} &\text{ if } v_{i,j}=1,\\
    1-\left(\frac{
    p}{1-p}\right)^{w_{i,j}} &\text{ if } v_{i,j}=0,
    \end{cases}    
\end{equation}
where $w_{i,j}$, $i \neq j$,  is the weight of the lowest-weight error that matches syndromes $i$ and $j$, whereas $w_{i,i}$ is the weight of the lowest-weight error that matches syndrome $i$ to the boundary. 

Overall, the bipartite graph $\mathcal{F}$ describes the decomposition of the probability distribution $\mathbb{P}_0$ into local factors defined over the configurations of  $v_{1,1}, \dots,  v_{s, s}$:
\begin{align}
\label{eq: prob fg}
   & \mathbb P_0(v_{1,1}, \dots, v_{1, s}, \dots, v_{s,1}, \dots, v_{s, s}) \propto  \nonumber \\
   & \qquad \prod_{1 \leq i \leq s} \mkern-10mu  \cnode_i(v_{i,1},\dots,v_{i,s} ) \mkern-15mu \prod_{1 \leq j \leq k \leq s} \mkern-15mu \prior_{jk} (v_{j,k}).  
\end{align}
We have omitted a normalization constant in \eqref{eq: prob fg}.






The factor nodes $\cnode_i$ strictly enforce the matching constraint by assigning a probability of zero to any invalid variable configuration. The remaining valid configurations, which correspond to perfect matchings, are then assigned probabilities proportional to the product of the independent prior probabilities $\prior_{i,j}$ of their constituent edges. Consequently, finding the minimum weight perfect matching is equivalent to identifying the configuration that maximizes the distribution $\mathbb{P}_0$.


\begin{figure*}
\centering
\begin{subfigure}[T]{0.4\textwidth}
  \scalebox{0.45}{
\begin{tikzpicture}[
    x=1.3cm, y=3.5cm,
    factor/.style={rectangle, draw=black!80, thick, minimum size=8mm, font=\normalsize},
    var/.style={circle, draw=black!80, thick, minimum size=5mm, font=\normalsize},
    edge/.style={thick, draw=black!70},
    prior/.style={->, >=stealth, thick, draw=black!70}
]
    \normalsize

    \def\dvar{1.35}   
    \def\dfac{1.90}   
    \def\xc{6.5}      

    \foreach \nm/\i/\lab/\p in {%
        11/1/{1,1}/{0.3}, 12/2/{1,2}/{0.2}, 13/3/{1,3}/{0.4}, 14/4/{1,4}/{0.9},
        22/5/{2,2}/{0.4}, 23/6/{2,3}/{0.3}, 24/7/{2,4}/{0.3},
        33/8/{3,3}/{0.8}, 34/9/{3,4}/{0.2},
        44/10/{4,4}/{0.4}%
    }{
        \pgfmathsetmacro{\xv}{\xc + (\i-5.5)*\dvar}
        \node[var] (v\nm) at (\xv, 0) {$\vnode_{\lab}$};
        \node[below=0.7cm of v\nm] (l\nm) {$\p$};
        \draw[prior] (l\nm) -- (v\nm);
    }

    \foreach \j in {1,2,3,4}{
        \pgfmathsetmacro{\xf}{\xc + (\j-2.5)*\dfac}
        \node[factor] (c\j) at (\xf, 1) {$\cnode_\j$};
    }

    \foreach \v in {11,12,13,14} { \draw[edge] (c1) -- (v\v); }
    \foreach \v in {12,22,23,24} { \draw[edge] (c2) -- (v\v); }
    \foreach \v in {13,23,33,34} { \draw[edge] (c3) -- (v\v); }
    \foreach \v in {14,24,34,44} { \draw[edge] (c4) -- (v\v); }

\end{tikzpicture}
}
  \caption{}
\end{subfigure}
\hfil
\begin{subfigure}[T]{0.4\textwidth}
  \scalebox{0.45}{
\begin{tikzpicture}[
    x=1.3cm, y=3.5cm,
    factor/.style={rectangle, draw=black!80, thick, minimum size=8mm, font=\normalsize},
    var/.style={circle, draw=black!80, thick, minimum size=5mm, font=\normalsize},
    factorforced/.style={rectangle, draw=black!80, fill=blue!30, thick, minimum size=8mm, font=\normalsize},
    varforced/.style={circle, draw=black!80, fill=blue!30, thick, minimum size=5mm, font=\normalsize},
    vargray/.style={circle, draw=black!20, fill=black!20, thick, minimum size=5mm, font=\normalsize},
    edge/.style={thick, draw=black!70},
    edgeforced/.style={very thick, draw=black},
    edgegray/.style={thick, draw=black!10},
    prior/.style={->, >=stealth, thick, draw=black!70},
    priorgray/.style={->, >=stealth, thick, draw=black!15}
]
    \normalsize

    \def\dvar{1.35}   
    \def\dfac{1.90}   
    \def\xc{6.5}      

    \foreach \nm/\i in {11/1, 12/2, 13/3, 14/4, 22/5, 23/6, 24/7, 33/8, 34/9, 44/10}{
        \pgfmathsetmacro{\xv}{\xc + (\i-5.5)*\dvar}
        \coordinate (p\nm) at (\xv, 0);
    }
    \foreach \j in {1,2,3,4}{
        \pgfmathsetmacro{\xf}{\xc + (\j-2.5)*\dfac}
        \coordinate (q\j) at (\xf, 1);
    }

    \node[vargray]   (v11) at (p11) {$\vnode_{1,1}$};
    \node[vargray]   (v12) at (p12) {$\vnode_{1,2}$};
    \node[vargray]   (v13) at (p13) {$\vnode_{1,3}$};
    \node[varforced] (v14) at (p14) {$\vnode_{1,4}$};

    \node[var]       (v22) at (p22) {$\vnode_{2,2}$};
    \node[var]       (v23) at (p23) {$\vnode_{2,3}$};
    \node[vargray]   (v24) at (p24) {$\vnode_{2,4}$};

    \node[var]       (v33) at (p33) {$\vnode_{3,3}$};
    \node[vargray]   (v34) at (p34) {$\vnode_{3,4}$};

    \node[vargray]   (v44) at (p44) {$\vnode_{4,4}$};

    \node[factorforced] (c1) at (q1) {$\cnode_1$};
    \node[factor]       (c2) at (q2) {$\cnode_2$};
    \node[factor]       (c3) at (q3) {$\cnode_3$};
    \node[factorforced] (c4) at (q4) {$\cnode_4$};

    \draw[edgegray]  (c1) -- (v11);
    \draw[edgegray]  (c1) -- (v12);
    \draw[edgegray]  (c1) -- (v13);
    \draw[edgeforced](c1) -- (v14);

    \draw[edgegray]  (c2) -- (v12);
    \draw[edge]      (c2) -- (v22);
    \draw[edge]      (c2) -- (v23);
    \draw[edgegray]  (c2) -- (v24);

    \foreach \v in {13, 23, 33, 34} { \draw[edge] (c3) -- (v\v); }

    \draw[edgeforced](c4) -- (v14);
    \draw[edgegray]  (c4) -- (v24);
    \draw[edgegray]  (c4) -- (v34);
    \draw[edgegray]  (c4) -- (v44);

    \node[below=0.7cm of v11] (l11) {\textcolor{black!20}{$0.3$}}; \draw[prior] (l11) -- (v11);
    \node[below=0.7cm of v12] (l12) {\textcolor{black!20}{$0.2$}}; \draw[prior] (l12) -- (v12);
    \node[below=0.7cm of v13] (l13) {\textcolor{black!20}{$0.4$}}; \draw[prior] (l13) -- (v13);
    \node[below=0.7cm of v14] (l14) {$\bf{0.9}$};                  \draw[prior] (l14) -- (v14);

    \node[below=0.7cm of v22] (l22) {$0.4$}; \draw[prior] (l22) -- (v22);
    \node[below=0.7cm of v23] (l23) {$0.3$}; \draw[prior] (l23) -- (v23);
    \node[below=0.7cm of v24] (l24) {$0.3$}; \draw[prior] (l24) -- (v24);

    \node[below=0.7cm of v33] (l33) {$0.8$}; \draw[prior] (l33) -- (v33);
    \node[below=0.7cm of v34] (l34) {$0.2$}; \draw[prior] (l34) -- (v34);

    \node[below=0.7cm of v44] (l44) {\textcolor{black!20}{$0.4$}}; \draw[prior] (l44) -- (v44);

\end{tikzpicture}
}
  \caption{}
\end{subfigure}
\hfil

\begin{subfigure}[T]{0.4\textwidth}
  \scalebox{0.45}{
\begin{tikzpicture}[
    x=1.3cm, y=3.5cm,
    factor/.style={rectangle, draw=black!80, thick, minimum size=8mm, font=\normalsize},
    var/.style={circle, draw=black!80, thick, minimum size=5mm, font=\normalsize},
    factorforced/.style={rectangle, draw=black!80, fill=blue!30, thick, minimum size=8mm, font=\normalsize},
    varforced/.style={circle, draw=black!80, fill=blue!30, thick, minimum size=5mm, font=\normalsize},
    vargray/.style={circle, draw=black!20, fill=black!20, thick, minimum size=5mm, font=\normalsize},
    edge/.style={thick, draw=black!70},
    edgeforced/.style={very thick, draw=black},
    edgegray/.style={thick, draw=black!10},
    prior/.style={->, >=stealth, thick, draw=black!70},
    priorgray/.style={->, >=stealth, thick, draw=black!15}
]
    \normalsize

    \def\dvar{1.35}   
    \def\dfac{1.90}   
    \def\xc{6.5}      

    \foreach \nm/\i in {11/1, 12/2, 13/3, 14/4, 22/5, 23/6, 24/7, 33/8, 34/9, 44/10}{
        \pgfmathsetmacro{\xv}{\xc + (\i-5.5)*\dvar}
        \coordinate (p\nm) at (\xv, 0);
    }
    \foreach \j in {1,2,3,4}{
        \pgfmathsetmacro{\xf}{\xc + (\j-2.5)*\dfac}
        \coordinate (q\j) at (\xf, 1);
    }

    \node[vargray]   (v11) at (p11) {$\vnode_{1,1}$};
    \node[vargray]   (v12) at (p12) {$\vnode_{1,2}$};
    \node[vargray]   (v13) at (p13) {$\vnode_{1,3}$};
    \node[varforced] (v14) at (p14) {$\vnode_{1,4}$};

    \node[var]       (v22) at (p22) {$\vnode_{2,2}$};
    \node[vargray]   (v23) at (p23) {$\vnode_{2,3}$};
    \node[vargray]   (v24) at (p24) {$\vnode_{2,4}$};

    \node[varforced] (v33) at (p33) {$\vnode_{3,3}$};
    \node[vargray]   (v34) at (p34) {$\vnode_{3,4}$};

    \node[vargray]   (v44) at (p44) {$\vnode_{4,4}$};

    \node[factorforced] (c1) at (q1) {$\cnode_1$};
    \node[factor]       (c2) at (q2) {$\cnode_2$};
    \node[factorforced] (c3) at (q3) {$\cnode_3$};
    \node[factorforced] (c4) at (q4) {$\cnode_4$};

    \draw[edgegray]   (c1) -- (v11);
    \draw[edgegray]   (c1) -- (v12);
    \draw[edgegray]   (c1) -- (v13);
    \draw[edgeforced] (c1) -- (v14);

    \draw[edgegray]   (c2) -- (v12);
    \draw[edge]       (c2) -- (v22);
    \draw[edgegray]   (c2) -- (v23);
    \draw[edgegray]   (c2) -- (v24);

    \draw[edgegray]   (c3) -- (v13);
    \draw[edgegray]   (c3) -- (v23);
    \draw[edgeforced] (c3) -- (v33);
    \draw[edgegray]   (c3) -- (v34);

    \draw[edgeforced] (c4) -- (v14);
    \draw[edgegray]   (c4) -- (v24);
    \draw[edgegray]   (c4) -- (v34);
    \draw[edgegray]   (c4) -- (v44);

    \node[below=0.7cm of v11] (l11) {\textcolor{black!20}{$0.3$}}; \draw[prior] (l11) -- (v11);
    \node[below=0.7cm of v12] (l12) {\textcolor{black!20}{$0.2$}}; \draw[prior] (l12) -- (v12);
    \node[below=0.7cm of v13] (l13) {\textcolor{black!20}{$0.4$}}; \draw[prior] (l13) -- (v13);
    \node[below=0.7cm of v14] (l14) {$\bf{0.9}$};                  \draw[prior] (l14) -- (v14);

    \node[below=0.7cm of v22] (l22) {$0.4$}; \draw[prior] (l22) -- (v22);
    \node[below=0.7cm of v23] (l23) {$0.3$}; \draw[prior] (l23) -- (v23);
    \node[below=0.7cm of v24] (l24) {$0.3$}; \draw[prior] (l24) -- (v24);

    \node[below=0.7cm of v33] (l33) {$\bf{0.8}$};                   \draw[prior] (l33) -- (v33);
    \node[below=0.7cm of v34] (l34) {\textcolor{black!20}{$0.2$}}; \draw[prior] (l34) -- (v34);

    \node[below=0.7cm of v44] (l44) {\textcolor{black!20}{$0.4$}}; \draw[prior] (l44) -- (v44);

\end{tikzpicture}
}
  \caption{}
\end{subfigure}
\hfil
\begin{subfigure}[T]{0.4\textwidth}
  \scalebox{0.45}{
\begin{tikzpicture}[
    x=1.3cm, y=3.5cm,
    factor/.style={rectangle, draw=black!80, thick, minimum size=8mm, font=\normalsize},
    var/.style={circle, draw=black!80, thick, minimum size=5mm, font=\normalsize},
    factorforced/.style={rectangle, draw=black!80, fill=blue!30, thick, minimum size=8mm, font=\normalsize},
    varforced/.style={circle, draw=black!80, fill=blue!30, thick, minimum size=5mm, font=\normalsize},
    vargray/.style={circle, draw=black!20, fill=black!20, thick, minimum size=5mm, font=\normalsize},
    edge/.style={thick, draw=black!70},
    edgeforced/.style={very thick, draw=black},
    edgegray/.style={thick, draw=black!10},
    prior/.style={->, >=stealth, thick, draw=black!70},
    priorgray/.style={->, >=stealth, thick, draw=black!15}
]
    \normalsize

    \def\dvar{1.35}   
    \def\dfac{1.90}   
    \def\xc{6.5}      

    \foreach \nm/\i in {11/1, 12/2, 13/3, 14/4, 22/5, 23/6, 24/7, 33/8, 34/9, 44/10}{
        \pgfmathsetmacro{\xv}{\xc + (\i-5.5)*\dvar}
        \coordinate (p\nm) at (\xv, 0);
    }
    \foreach \j in {1,2,3,4}{
        \pgfmathsetmacro{\xf}{\xc + (\j-2.5)*\dfac}
        \coordinate (q\j) at (\xf, 1);
    }

    \node[vargray]   (v11) at (p11) {$\vnode_{1,1}$};
    \node[vargray]   (v12) at (p12) {$\vnode_{1,2}$};
    \node[vargray]   (v13) at (p13) {$\vnode_{1,3}$};
    \node[varforced] (v14) at (p14) {$\vnode_{1,4}$};

    \node[varforced] (v22) at (p22) {$\vnode_{2,2}$};
    \node[vargray]   (v23) at (p23) {$\vnode_{2,3}$};
    \node[vargray]   (v24) at (p24) {$\vnode_{2,4}$};

    \node[varforced] (v33) at (p33) {$\vnode_{3,3}$};
    \node[vargray]   (v34) at (p34) {$\vnode_{3,4}$};

    \node[vargray]   (v44) at (p44) {$\vnode_{4,4}$};

    \node[factorforced] (c1) at (q1) {$\cnode_1$};
    \node[factorforced] (c2) at (q2) {$\cnode_2$};
    \node[factorforced] (c3) at (q3) {$\cnode_3$};
    \node[factorforced] (c4) at (q4) {$\cnode_4$};

    \draw[edgegray]   (c1) -- (v11);
    \draw[edgegray]   (c1) -- (v12);
    \draw[edgegray]   (c1) -- (v13);
    \draw[edgeforced] (c1) -- (v14);

    \draw[edgegray]   (c2) -- (v12);
    \draw[edgeforced] (c2) -- (v22);
    \draw[edgegray]   (c2) -- (v23);
    \draw[edgegray]   (c2) -- (v24);

    \draw[edgegray]   (c3) -- (v13);
    \draw[edgegray]   (c3) -- (v23);
    \draw[edgeforced] (c3) -- (v33);
    \draw[edgegray]   (c3) -- (v34);

    \draw[edgeforced] (c4) -- (v14);
    \draw[edgegray]   (c4) -- (v24);
    \draw[edgegray]   (c4) -- (v34);
    \draw[edgegray]   (c4) -- (v44);

    \node[below=0.7cm of v11] (l11) {\textcolor{black!20}{$0.3$}}; \draw[prior] (l11) -- (v11);
    \node[below=0.7cm of v12] (l12) {\textcolor{black!20}{$0.2$}}; \draw[prior] (l12) -- (v12);
    \node[below=0.7cm of v13] (l13) {\textcolor{black!20}{$0.4$}}; \draw[prior] (l13) -- (v13);
    \node[below=0.7cm of v14] (l14) {$\bf{0.9}$};                  \draw[prior] (l14) -- (v14);

    \node[below=0.7cm of v22] (l22) {$\bf{0.4}$};                  \draw[prior] (l22) -- (v22);
    \node[below=0.7cm of v23] (l23) {\textcolor{black!20}{$0.3$}}; \draw[prior] (l23) -- (v23);
    \node[below=0.7cm of v24] (l24) {\textcolor{black!20}{$0.3$}}; \draw[prior] (l24) -- (v24);

    \node[below=0.7cm of v33] (l33) {$\bf{0.8}$};                  \draw[prior] (l33) -- (v33);
    \node[below=0.7cm of v34] (l34) {\textcolor{black!20}{$0.2$}}; \draw[prior] (l34) -- (v34);

    \node[below=0.7cm of v44] (l44) {\textcolor{black!20}{$0.4$}}; \draw[prior] (l44) -- (v44);

\end{tikzpicture}
}
  \caption{}
\end{subfigure}
\caption{Marginalization and forced convergence: (a) An example marginalization output for the error candidate in Fig.~\ref{fig: different factor graphs}. We include the (probability) marginalization result after $T$ iterations $\mathbb P^{(T)}(v=1)$ for each variable $\vnode \in \setvnodes$. (b) The first decision made by forced convergence. We color in blue the chosen variable together with the unsatisfied syndromes it matches. We color in gray the variables and the edges that are eliminated for future steps because of the choice. (c)  The second decision made by forced convergence. This decision, together with the previous one constitute the output of marginalization. Marginalization is not guaranteed to converge and, in fact, in this example it does not. (d) The third decision made by forced convergence. This decision, together with the previous ones, constitute the output of forced convergence. Forced convergence is guaranteed to converge and, in fact, in this example it does.}
\label{fig: inference example}
\label{fig: marg and forced conv}
\end{figure*}

\subsection{Belief Propagation}

Maximizing $\mathbb{P}_0$ is a computationally demanding task. Instead, in this paper, we aim at a low-complexity approximation of this solution. In particular, we apply the sum-product belief propagation algorithm over the factor graph $\mf$. This yields the marginal probability of each variable in $\mf$, which represents the likelihood of that variable being present in a valid matching, weighted by the overall likelihood of the matchings it belongs to. We then rely on these marginals to heuristically approximate (see Section \ref{sec: forced conv}) the specific configuration that maximizes $\mathbb{P}_0$.

In the sum-product algorithm, factor and variable nodes exchange messages over the edges $\setedges$ of the bipartite graph $\graph$.For numerical stability and notational convenience, we provide a log-domain implementation of the sum-product algorithm.
The log-likelihood ratio associated to the prior probability
distribution~\eqref{eq: two-syn var factors} of a variable node $\vnode \in \setvnodes$
is denoted by $\priorlog_{\vnode}$. For $\vnode = \vnode_{i,j}$,
\begin{equation}
    \priorlog_{\vnode_{i,j}}
    = \ln \frac{\rho_{i,j}(v_{i,j}=1)}{\rho_{i,j}(v_{i,j}=0)}
    \label{eq:priorllr}
\end{equation}

The message sent by factor node $\cnode$ to variable node $\vnode$ at the $t$-th iteration is denoted by $\mcv^{(t)}$, whereas the message sent from $\vnode$ to $\cnode$ is denoted by $\mvc^{(t)}$.
Initially, the messages from variables to factors $\mvc^{(0)}$ are
initialized to $\priorlog_{\vnode}$, whereas the messages from factors
to variables $\mcv^{(0)}$ are initialized to zero.
%
In the subsequent iterations, the messages from variable nodes to factor nodes are updated as
$$ 
\mvc^{(t+1)} = \priorlog_{\vnode} + \sum_{\cnode' \in \nn(\vnode) \smin \cnode} \mkern-20mu m_{\cnode' \to \vnode}^{(t)} 
$$
whereas the messages from factor to variable nodes are updated as
$$ 
\mcv^{(t+1)} = - {\max}^*_{\vnode' \in \nn(\cnode) \smin \vnode} \left( m_{\vnode' \to \cnode}^{(t)} \right)
$$
where $\max^*$ is the soft-max operator, defined over a set of variables $x_1, \dots, x_n$ as:
$$ 
{\max}^*(x_1, \dots, x_n) = \ln \left( \sum_{i=1}^n \exp(x_i) \right) 
$$
In practice, computing the multi-argument soft-max operator directly via exponentials can lead to numerical instability. Therefore, the soft-max operation is typically evaluated recursively using its two-argument form, sometimes known as the Jacobian logarithm:
$$
{\max}^* (x, y) \approx \max(x, y) + \ln(1 + \exp(-|x - y|)).
$$

Similarly, the posterior probability of the random variable associated with $\vnode$ after the $t$-th iteration is obtained as
\begin{equation}
\posteriorlog_{\vnode}^{(t+1)} = \priorlog_{\vnode} + \sum_{\cnode \in \nn(\vnode) } \text{ }\mkern-20mu m_{\cnode \to \vnode}^{(t)}.
\label{eq: marginalization}
\end{equation}
for a predefined maximum number of  iterations $T$.

Each variable node is connected to at most two factor nodes, so each of
its outgoing messages can be updated in $O(1)$ time. There are $O(n^2)$
variable nodes, and the update of one is independent of the others. With
$O(n^2)$ parallel resources, all of them can therefore be updated
simultaneously in $O(1)$ time. With only $O(n)$ resources, each of them
handles $O(n)$ variable nodes serially, and the update of all variable
nodes takes $O(n)$ time.
In contrast, the factors are adjacent to $O(n)$ variable nodes. Using the leave-one-out approach~\cite{mezard2009information}, all messages outgoing from a factor node can be computed in $O(n)$ time in total. Since all factor nodes can be updated in parallel (i.e. we assume we have access to $O(n)$ parallel resources), each iteration over them takes time $O(n)$. Lastly, saving the messages requires $O(n^2)$ space. Overall, each iteration can be done in $O(n)$ time and requires $O(n^2)$ space.

\subsection{Checking convergence}
\label{sec:check_conv}


Checking convergence in our algorithms is useful since there is no guarantee that message-passing will achieve it. Hence, if we check for it early, it can save us computational time.

Assuming that we have inferred an error candidate, checking for convergence requires for each factor node to sum over the inferred values of the variable nodes adjacent to it. If the sum is equal to one for all factor nodes, then we have converged. This can be parallelized for each factor node and, as a result, it requires $O(n)$ time.

\begin{table*}[!t]
\centering
\begin{tabular}{|l | c c c c | c | c |}
\hline
\textbf{Algorithm} &
\textbf{Noise mod.} &
\textbf{Iter. Comp.} &
\textbf{\#Iter. ($T$) } &
\textbf{Inf. Comp.} &
\textbf{Overall Comp.} &
\textbf{Threshold} \\
\hline
BP4M  & Capacity & $O(n)$ & $25$  & $O(n \log n)$  & $O(Tn + n \log n)$ & 0.1170\\
BP4MF & Capacity  & $O(n \log n)$ & $25$ & -  & $O(T n \log n)$ & 0.1340\\
BP4M+M  & Capacity  & $O(n)$ & $25$  & $O(r_{nc} n^3 \log n)$  & $O((1-r_{nc}) T n + r_{nc} n^3 \log n)$ & 0.1570\\
B-BP4MF  & Capacity  & $O(n \log n)$ & $25+25$  & -  & $O(T n \log n)$ & 0.1500\\
MWPM  & Capacity  & - & -  & -  & $O(n^3 \log n)$ & 0.1550\\
BP4M  & Circuit-level & $O(n)$ & $50$  & $O(n \log n)$  & $O(Tn + n \log n)$ & 0.0050\\
BP4MF & Circuit-level  & $O(n \log n)$ & $50$ & -  & $O(T n \log n)$ & 0.0055\\
B-BP4MF  & Circuit-level  & $O(n \log n)$ & $75+75$  & -  & $O(T n \log n)$ & 0.0061\\
MWPM  & Circuit-level  & - & -  & -  & $O(n^3 \log n)$ & 0.0077\\
\hline
\end{tabular}
\caption{Algorithms' worst case complexity and thresholds for the unrotated surface code. The second column is the noise model (circuit-level stands for memory-Z experiment); the third column is the complexity per iteration; the fourth column is the number of iterations;
the fifth column is the complexity of inference after BP (a hyphen indicates that this step is not needed since BP always converges); the sixth column is the overall complexity; the seventh column is the observed threshold. We have taken the MWPM threshold from the literature~\cite{wang2009threshold} for code capacity, and we have simulated it through Pymatching~\cite{higgott2022pymatching} for circuit-level noise. 
The complexities reported here are for a parallel implementation (see Sections \ref{sec:methods} and \ref{sec:res}).}
\label{table: complexities}
\end{table*}

\subsection{Forced convergence}
\label{sec: forced conv}

We infer an error candidate $\bm{e'}$ by using the set of messages from checks to variables. We consider two inference methods: \textbf{marginalization} and \textbf{forced convergence}. 

In marginalization, we infer the posterior probability distribution
$\posteriorlog_{\vnode}^{(t)}$
for each individual variable $\vnode \in \setvnodes$
and include $\vnode$ in the matching provided $\posteriorlog_{\vnode}^{(t)}>0$.

The marginalization over each variable requires $O(1)$ time since each variable node is only connected to at most two factor nodes. Since we can marginalize each variable in parallel, we only require $O(1)$ time for marginalizing provided we use $O(n^2)$ space (and, as in the case of variable to factor updates, $O(n)$ time if we only use $O(n)$ space).
Fig.~\ref{fig: inference example} is an example where marginalization does not converge.

Given that marginalization decides locally on each variable,
we are not guaranteed to converge. Fig.~\ref{fig: calls matching} provides data for the number of times we do not obtain (through marginalization) a converged error candidate in any of the $T=25$ iteration steps.

\begin{figure*}[t]
\centering
\begin{minipage}{0.49\textwidth}\centering
  \textbf{BP4M}\\[0.5em]
  \resizebox{\linewidth}{!}{\input{plots/plot_tex/threshold_bp4m_25}}
\end{minipage}\hfill
\begin{minipage}{0.49\textwidth}\centering
  \textbf{BP4MF}\\[0.5em]
  \resizebox{\linewidth}{!}{\input{plots/plot_tex/threshold_bp4mf_25_iter}}
\end{minipage}
\caption{Code capacity unrotated surface code. We ran our algorithms for 25 iterations.
The horizontal red dotted line is the \textbf{pseudo-threshold} (i.e. the curve where the logical error rate equals the physical error rate) and the vertical dotted line indicates the achieved threshold.
All subfigures include a gray LER curve that represents pure MWPM distance 11 decoding performance. This is for comparison reasons with the corresponding distance 11 LER curve of our methods (orange).} 
\label{fig: threshold plots}
\end{figure*}

\begin{figure*}
\centering
\begin{minipage}[t]{0.49\textwidth}\centering
  \textbf{ BP4M}\\[0.5em]
  \resizebox{\linewidth}{!}{\input{plots/plot_tex/threshdol_bp4m_rotated_memory_z}}
\end{minipage}\hfill
\begin{minipage}[t]{0.49\textwidth}\centering
  \textbf{ BP4MF}\\[0.5em]
  \resizebox{\linewidth}{!}{
\begin{tikzpicture}

\newcommand{\cpms}{4.5pt}

\pgfplotsset{
  cpcurve/.style={
    color=#1,
    line width=3.5pt,
    mark=*,
    mark size=\cpms,
    mark options={solid, fill=#1, draw=#1, line width=1.2pt},
    error bars/y dir=both,
    error bars/y explicit,
    error bars/error bar style={color=#1, line width=2.5pt},
    error bars/error mark=-,
    error bars/error mark options={color=#1, mark size=3pt, line width=2.5pt},
  },
  cpkey/.style={
    color=#1, line width=3.5pt, mark=*, mark size=\cpms,
    mark options={solid, fill=#1, draw=#1, line width=1.2pt},
  },
}

\begin{axis}[
  xmode=log, ymode=log,
  axis background/.style={fill=white},
  width=22.86cm, height=15.24cm,         
  scale only axis=false,
  xmin=0.0018286156535236553, xmax=0.01312468257271731,
  ymin=8.474540288069868e-05, ymax=0.49695778975115007,
  label style={font=\fontsize{21}{25}\selectfont},
  tick label style={font=\fontsize{22}{26}\selectfont},
  xtick={0.002,0.003,0.004,0.006,0.01},
  xticklabels={$2\times10^{-3}$,$3\times10^{-3}$,$4\times10^{-3}$,
               $6\times10^{-3}$,$10^{-2}$},
  minor xtick={0.005,0.007,0.008,0.009},
  ytick={1e-4,1e-3,1e-2,1e-1},
  minor ytick={9e-05,0.0002,0.0003,0.0004,0.0005,0.0006,0.0007,0.0008,
              0.0009,0.002,0.003,0.004,0.005,0.006,0.007,0.008,0.009,
              0.02,0.03,0.04,0.05,0.06,0.07,0.08,0.09,0.2,0.3,0.4},
  tick align=outside,
  xtick pos=left, ytick pos=left,
  grid=both,
  grid style      ={mplgrid, dotted, line width=1pt},
  minor grid style={mplgrid, dotted, line width=1pt},
  axis on top=false,
]

  \addplot[forget plot, mplbrown, mpldotted, line width=3.5pt]
    coordinates {(0.0055,8.474540288069868e-05) (0.0055,0.49695778975115007)};
  \addplot[forget plot, black,    mpldotted, line width=3.5pt]
    coordinates {(0.0072,8.474540288069868e-05) (0.0072,0.49695778975115007)};

  \addplot[cpcurve=mplblue, forget plot]
    table[x=p, y=ler, y error plus=eplus, y error minus=eminus] {
      p        ler              eplus            eminus
      0.002    0.003088888889   0.0003063390912  0.0002857464471
      0.003    0.0062625        0.0006546616967  0.0006085929494
      0.004    0.01116666667    0.001000473583   0.0009397660791
      0.005    0.01735          0.001521099615   0.001431206341
      0.006    0.0264           0.002654980367   0.002478269765
      0.007    0.03235          0.00291683701    0.002742559384
      0.008    0.0402           0.003224211221   0.003053060035
      0.009    0.05265          0.003648076127   0.00348177338
      0.01     0.05925          0.003848953755   0.003685187932
      0.011    0.06955          0.00413686961    0.003977033465
      0.012    0.08305          0.004476102533   0.004321379594
    };

  \addplot[cpcurve=mplgreen, forget plot]
    table[x=p, y=ler, y error plus=eplus, y error minus=eminus] {
      p        ler              eplus            eminus
      0.002    0.001177272727   0.0001212810672  0.0001128206379
      0.003    0.00365          0.0003781845153  0.0003517238632
      0.004    0.008333333333   0.0008710463469  0.0008098777742
      0.005    0.016125         0.001469525227   0.001379366407
      0.006    0.02765          0.002712483325   0.002536290035
      0.007    0.041            0.003253557054   0.003082720766
      0.008    0.05705          0.003783559592   0.003618950059
      0.009    0.0738           0.004247897245   0.004089674636
      0.01     0.09675          0.0047850315     0.00463546857
      0.011    0.1235           0.005309159239   0.005169619141
      0.012    0.15005          0.005749639548   0.005620005279
    };

  \addplot[cpcurve=mplred, forget plot]
    table[x=p, y=ler, y error plus=eplus, y error minus=eminus] {
      p        ler              eplus            eminus
      0.002    0.0005644444444  5.876386668e-05  5.46220584e-05
      0.003    0.002431818182   0.0002461606131  0.0002292857428
      0.004    0.007325         0.0007051631882  0.0006592390099
      0.005    0.0163           0.001477016316   0.0013868957
      0.006    0.0285           0.002750779072   0.002574935467
      0.007    0.0492           0.003537067329   0.003369430653
      0.008    0.07385          0.004249178993   0.004090975344
      0.009    0.10325          0.004921161862   0.004774039727
      0.01     0.13775          0.005554125623   0.005419906124
      0.011    0.1766           0.006128325636   0.006008571637
      0.012    0.2092           0.006524888012   0.006417243145
    };

  \addplot[cpcurve=mplpurple, forget plot]
    table[x=p, y=ler, y error plus=eplus, y error minus=eminus] {
      p        ler              eplus            eminus
      0.002    0.000286         4.029883099e-05  3.65529768e-05
      0.003    0.00174          0.0001784777828  0.0001660837454
      0.004    0.00592          0.0005673286227  0.0005304869542
      0.005    0.0156           0.001446792051   0.001356518035
      0.006    0.03235          0.00291683701    0.002742559384
      0.007    0.05575          0.003744201102   0.003579092147
      0.008    0.08875          0.004608520946   0.004455947778
      0.009    0.1319           0.005456094516   0.005319691992
      0.01     0.17395          0.006092952732   0.005972213434
      0.011    0.22425          0.006686105797   0.006584045191
      0.012    0.2737           0.007132833869   0.007049103247
    };

  \addplot[cpcurve=mplorange, forget plot]
    table[x=p, y=ler, y error plus=eplus, y error minus=eminus] {
      p        ler              eplus            eminus
      0.002    0.000152         3.005249895e-05  2.628181012e-05
      0.003    0.001070833333   0.0001107702789  0.0001030128434
      0.004    0.004541666667   0.000454775299   0.000423972633
      0.005    0.0133           0.001342130272   0.001251339334
      0.006    0.0336           0.002968379798   0.002794605239
      0.007    0.06095          0.003898484356   0.003735369317
      0.008    0.10535          0.004963848292   0.004817513886
      0.009    0.1588           0.005880627309   0.005754251548
      0.01     0.21495          0.006588001992   0.006482491026
      0.011    0.2802           0.007182972725   0.007101649891
      0.012    0.3275           0.007494388234   0.007430578735
    };

  \addplot[forget plot, mplbrown, mpldashdot, line width=3.5pt,
           domain=0.0018286156535236553:0.01312468257271731, samples=2] {x};

  \node[anchor=north west,
        draw=black!25, fill=white, fill opacity=0.9, text opacity=1,
        rounded corners=2pt, inner sep=6pt,
        font=\fontsize{18}{22}\selectfont]
    at (rel axis cs:0.02,0.98) {Rotated Surface memory-Z};

  \node[anchor=south east, mplbrown,
        font=\bfseries\fontsize{20}{24}\selectfont]
    at (axis cs:0.00520, 9.3e-4) {BP4MF};
  \node[anchor=south east, mplbrown,
        font=\bfseries\fontsize{20}{24}\selectfont]
    at (axis cs:0.00520, 1.83e-3) {0.0055};
  \node[anchor=south west, black,
        font=\bfseries\fontsize{20}{24}\selectfont]
    at (axis cs:0.00755, 9.3e-4) {MWPM};
  \node[anchor=south west, black,
        font=\bfseries\fontsize{20}{24}\selectfont]
    at (axis cs:0.00755, 1.83e-3) {0.0072};

\end{axis}
\end{tikzpicture}}
\end{minipage}

\vspace{1.5em}

\begin{minipage}[t]{0.49\textwidth}\centering
  \resizebox{\linewidth}{!}{
\begin{tikzpicture}

\newcommand{\cpms}{4.5pt}

\pgfplotsset{
  cpcurve/.style={
    color=#1,
    line width=3.5pt,
    mark=*,
    mark size=\cpms,
    mark options={solid, fill=#1, draw=#1, line width=1.2pt},
    error bars/y dir=both,
    error bars/y explicit,
    error bars/error bar style={color=#1, line width=2.5pt},
    error bars/error mark=-,
    error bars/error mark options={color=#1, mark size=3pt, line width=2.5pt},
  },
  cpkey/.style={
    color=#1, line width=3.5pt, mark=*, mark size=\cpms,
    mark options={solid, fill=#1, draw=#1, line width=1.2pt},
  },
}

\begin{axis}[
  xmode=log, ymode=log,
  axis background/.style={fill=white},
  width=22.86cm, height=15.8cm,         
  scale only axis=false,
  xmin=0.0018286156535236553, xmax=0.01312468257271731,
  ymin=0.00015434979337183868, ymax=0.6882253121999693,
  xlabel={physical error rate },
  ylabel={LER},
  xlabel style={font=\fontsize{23}{27}\selectfont},
  ylabel style={font=\fontsize{30}{34}\selectfont},
  tick label style={font=\fontsize{22}{26}\selectfont},
  xtick={0.002,0.003,0.004,0.006,0.01},
  xticklabels={$2\times10^{-3}$,$3\times10^{-3}$,$4\times10^{-3}$,
               $6\times10^{-3}$,$10^{-2}$},
  minor xtick={0.005,0.007,0.008,0.009},
  ytick={1e-3,1e-2,1e-1},
  minor ytick={0.0002,0.0003,0.0004,0.0005,0.0006,0.0007,0.0008,0.0009,
              0.002,0.003,0.004,0.005,0.006,0.007,0.008,0.009,0.2,0.3,0.4,0.5,0.6},
  tick align=outside,
  xtick pos=left, ytick pos=left,
  grid=both,
  grid style      ={mplgrid, dotted, line width=1pt},
  minor grid style={mplgrid, dotted, line width=1pt},
  axis on top=false,
]

  \addplot[forget plot, mplbrown, mpldotted, line width=3.5pt]
    coordinates {(0.005,0.00015434979337183868) (0.005,0.6882253121999693)};
  \addplot[forget plot, black,    mpldotted, line width=3.5pt]
    coordinates {(0.0077,0.00015434979337183868) (0.0077,0.6882253121999693)};

  \addplot[cpcurve=mplblue, forget plot]
    table[x=p, y=ler, y error plus=eplus, y error minus=eminus] {
      p        ler              eplus            eminus
      0.002    0.004508928571   0.0004698502089  0.0004368289953
      0.003    0.00925          0.0009486676154  0.0008832669463
      0.004    0.01476470588    0.001534547268   0.001428015319
      0.005    0.02363636364    0.002399810523   0.002238212726
      0.006    0.034875         0.003391327537   0.003174461558
      0.007    0.04741666667    0.004531626405   0.004250290252
      0.008    0.0584           0.005471380389   0.005142009522
      0.009    0.06975          0.006646999831   0.006245738117
      0.01     0.081375         0.007109346423   0.006719271583
      0.011    0.1005           0.00903156992    0.008534967528
      0.012    0.1161666667     0.009592994436   0.009116239176
    };

  \addplot[cpcurve=mplgreen, forget plot]
    table[x=p, y=ler, y error plus=eplus, y error minus=eminus] {
      p        ler              eplus            eminus
      0.002    0.001407303371   0.0001475116051  0.0001370574345
      0.003    0.004901960784   0.0005133588515  0.000477126614
      0.004    0.01134782609    0.001157796543   0.001078522087
      0.005    0.02313636364    0.002376088975   0.002214299396
      0.006    0.03785714286    0.003775011136   0.003528670721
      0.007    0.0579           0.005450428355   0.005120667806
      0.008    0.082            0.007132953687   0.00674347758
      0.009    0.1111666667     0.00942016482    0.008937087879
      0.01     0.15             0.01310337746    0.01245060766
      0.011    0.174            0.0138610609     0.01325351345
      0.012    0.205            0.01470273059    0.0141533683
    };

  \addplot[cpcurve=mplred, forget plot]
    table[x=p, y=ler, y error plus=eplus, y error minus=eminus] {
      p        ler              eplus            eminus
      0.002    0.000664         8.643949389e-05  7.896271089e-05
      0.003    0.003623188406   0.0003797084514  0.0003528590395
      0.004    0.01012          0.001050175985   0.0009770457398
      0.005    0.02363636364    0.002399810523   0.002238212726
      0.006    0.04508333333    0.004429298421   0.004146424859
      0.007    0.07925          0.007028180325   0.00663606793
      0.008    0.1151666667     0.00955887753    0.009080858774
      0.009    0.1635           0.01354207196    0.01291476256
      0.01     0.2095           0.01481360006    0.01427266789
      0.011    0.263            0.02274373674    0.02185839462
      0.012    0.2925           0.02342141646    0.02264664654
    };

  \addplot[cpcurve=mplpurple, forget plot]
    table[x=p, y=ler, y error plus=eplus, y error minus=eminus] {
      p        ler              eplus            eminus
      0.002    0.000431         4.891005191e-05  4.517726221e-05
      0.003    0.002657894737   0.0002772485867  0.0002577108466
      0.004    0.009481481481   0.0009782200492  0.0009104239517
      0.005    0.02322727273    0.002380422352   0.002218667699
      0.006    0.049            0.004599401592   0.004319104586
      0.007    0.08883333333    0.008571676623   0.008060204678
      0.008    0.15375          0.01322862803    0.01258293709
      0.009    0.201            0.0146018965     0.01404503773
      0.01     0.2715           0.02295150867    0.02209804698
      0.011    0.341            0.02428818463    0.02369485106
      0.012    0.3715           0.02468960412    0.02421021207
    };

  \addplot[cpcurve=mplorange, forget plot]
    table[x=p, y=ler, y error plus=eplus, y error minus=eminus] {
      p        ler              eplus            eminus
      0.002    0.000261         3.860774904e-05  3.485869326e-05
      0.003    0.0019609375     0.0002052656154  0.0001907440924
      0.004    0.00814516129    0.0008470484503  0.0007878343044
      0.005    0.0267           0.00266891397    0.002492327875
      0.006    0.0535           0.005261163212   0.004927961547
      0.007    0.1061666667     0.009241521277   0.008752111438
      0.008    0.18025          0.01404235343    0.01344655457
      0.009    0.23825          0.01546164958    0.01497450257
      0.01     0.3155           0.023868954      0.02318027172
      0.011    0.3795           0.02477759604    0.02432807532
      0.012    0.4445           0.0252391921     0.02503221719
    };

  \addplot[forget plot, mplbrown, mpldashdot, line width=3.5pt,
           domain=0.0018286156535236553:0.01312468257271731, samples=2] {x};

  \node[anchor=north west,
        draw=black!25, fill=white, fill opacity=0.9, text opacity=1,
        rounded corners=2pt, inner sep=6pt,
        font=\fontsize{20}{24}\selectfont]
    at (rel axis cs:0.02,0.98) {Unrotated Surface memory-Z};

  \node[anchor=south west, mplbrown,
        font=\bfseries\fontsize{20}{24}\selectfont]
    at (axis cs:0.00525, 1.62e-3) {BP4M};
  \node[anchor=south west, mplbrown,
        font=\bfseries\fontsize{20}{24}\selectfont]
    at (axis cs:0.00525, 2.69e-3) {0.005};
  \node[anchor=south west, black,
        font=\bfseries\fontsize{20}{24}\selectfont]
    at (axis cs:0.00790, 1.62e-3) {MWPM};
  \node[anchor=south west, black,
        font=\bfseries\fontsize{20}{24}\selectfont]
    at (axis cs:0.00790, 2.69e-3) {0.0077};

\end{axis}
\end{tikzpicture}}
\end{minipage}\hfill
\begin{minipage}[t]{0.49\textwidth}\centering
  \resizebox{\linewidth}{!}{\input{plots/plot_tex/threshold_bp4mf_unrotated_memory_z}}
\end{minipage}
\caption{Circuit-level noise memory-Z surface code (rotated and unrotated). We ran our algorithms for 50 iterations. The columns determine the algorithm, while the rows determine the code used.
The horizontal red dotted line is the pseudo-threshold and the vertical dotted line indicates the achieved threshold.}
\label{fig: threshold circuit level noise}

\end{figure*}


On the other hand, forced convergence uses the posteriors computed by marginalization for ordering all variable nodes in terms of their probability of being part of the matching. We then recursively pick the variable with the highest probability until we obtain a matching. The recursive decision on the variables together with their ordering in terms of likelihood provide global information that always allows us to provide an error candidate matching the syndrome. Forced convergence is somewhat reminiscent of \textbf{guided decimation}~\cite{yao2024belief}.

To force the convergence, we associate to each $\cnode \in \setcnodes$ 
a subset of its adjacent variable nodes
\begin{equation*}
    \me_\cnode \subseteq \nn(\cnode)
\end{equation*}
chosen so that that these subsets partition the variable nodes, 
\begin{equation*}
    \me = \cup_{\cnode \in \setcnodes} \me_\cnode \text{ and } \me_{\cnode} \cap \me_{\cnode'} = \emptyset.
\end{equation*}
This partition can be pre-computed assuming each syndrome is unsatisfied and obtaining a partition for the complete set of variable nodes that the decoding graph may be. Each $\me_{\cnode}$ can then be in parallel adapted to the observed syndrome for a total cost of $O(n)$.

Given the partition, we marginalize each variable in $O(1)$ and compute the block maxima
\begin{equation*}
\vnode^\cnode_{\text{max}} \equiv \text{arg max}_{\vnode \in \me_\cnode} \left\{\posteriorlog_{\vnode}^{(t)}\right\}   \end{equation*}
one for each $\cnode \in \setcnodes$. Since we can do this for each of them in parallel, this requires $O(n)$.

Next, we build the vector of maxima
\begin{equation*}
\bm{\vnode}_{\text{max}} = (\vnode^{\cnode_1}_{\text{max}},\dots,\vnode^{\cnode_s}_{\text{max}})    
\end{equation*}
and reorder it $\bm{\vnode}_{\text{max}}^{\pi}$ such that 
\begin{equation*}
\posteriorlog_{(\bm{\vnode}_{\text{max}}^\pi)_i=1}^{(t)}
\geq \posteriorlog_{(\bm{\vnode}_{\text{max}}^\pi)_{i+1}}^{(t)}
\end{equation*}
for $i=1,\dots,s-1$. When performed serially, reordering requires $O(n \log n)$, but there exist parallel hardware implementations~\cite{bascones2025hourglass} with even lower complexity.

We then process $\bm{\vnode}_{\text{max}}^{\pi}$ starting from its first entry, which carries
the largest posterior. At each step, we take the first remaining entry, say this entry is associated to check node $\cnode_i$, and proceed as follows.
If its representative is the boundary variable $\vnode_{i,i}$, we match
$\cnode_i$ to the boundary and remove the entry associated to $\cnode_i$
from $\bm{\vnode}_{\text{max}}^{\pi}$. Otherwise the representative is
$\vnode_{i,k}$ for some $k \neq i$, in which case we match $\cnode_i$ to
$\cnode_k$ and remove from $\bm{\vnode}_{\text{max}}^{\pi}$ both the entry
associated to $\cnode_i$ and the one associated to $\cnode_k$. 
We proceed in this way until every entry has been processed.

Since an entry is removed from $\bm{\vnode}_{\text{max}}^{\pi}$ exactly
when its check is matched, every check is matched exactly once, which yields a valid matching, i.e., an error candidate that reproduces the observed syndrome. 
Processing the entries costs $O(n)$, so forced convergence requires $O(n\log n)$ overall, dominated by the
reordering.

It is worth noting that, in case marginalization converges, running forced convergence with the same BP messages will provide the same output as marginalization. Hence, at a given iteration $t$, it is convenient to check first whether marginalization converged, at a cost of $O(n)$, in which case one can skip forced convergence, and save its $O(n\log n)$ cost. 
For illustration, Fig.~\ref{fig: inference example} is an example where marginalization does not converge and forced convergence does. 

\subsection{Correlated decoding}
\label{sec: correlated}

MWPM can approximate the joint decoding of $X$ and $Z$ Pauli errors by first running quaternary BP on the Tanner graph and then using the BP qubit posteriors in order to compute the input weights required by MWPM. This approach is known as Belief-matching \cite{higgott2023improved}. 

We can emulate Belief-matching to incorporate correlations into our scheme. In order to compute the prior probabilities that we associate with the edges in the decoding graph, we:
\begin{enumerate}
    \item Run quaternary BP on the full Tanner graph and compute its qubit posteriors.
    \item Run Dijkstra's algorithm \cite{dijkstra1959note} in order to get the path of minimum weight (according to the qubit BP posteriors) that joins either two pairs of unsatisfied syndromes or one unsatisfied syndrome to the boundary. 
\end{enumerate}

The graph $\mg_{\text{D}}=(\mv_{\text{D}},\me_{\text{D}})$ on which Dijkstra's algorithm is run is quite close to the Tanner graph: it has one node $i \in \mv_{\text{D}}$ for each check. This includes virtual checks, which are located at the check-free end of each qubit that is connected to a single check node. The graph has one edge $(i,j) \in \me_{\text{D}}$ for each qubit, with $i,j \in \mv_{\text{D}}$ being the two check nodes adjacent to that qubit. One of these checks might be virtual.
 
 


There is some subtlety regarding the graph on which we run Dijkstra, depending on whether we allow corrections between pairs of syndromes to go through the boundary or not. If we do, then we can use a quotient graph $\quot{\mg_{\text{D}}}=(\quot{\mv_{\text{D}}},\quot{\me_{\text{D}}})$, where $\quot{\mv_{\text{D}}}$ has one equivalence class for each non-virtual check and all the nodes associated to virtual checks are collapsed into a single equivalence class. Moreover, $i,j \in \quot{\mv_{\text{D}}}$ are connected by an edge in $\quot{\me_{\text{D}}}$ provided one member of the equivalence class of $i$ is connected to a member of that of $j$
in $\mv_{\text{D}}$.
We have given preference to $\mg_{\text{D}}$ over $\quot{\mg_{\text{D}}}$ in order to help BP on the decoding graph: If we take $\quot{\mg_{\text{D}}}$, we risk having some redundancy in the matchings. More specifically, a situation may arise where, given two unsatisfied syndromes, the solution that matches each of them separately to the boundary is equivalent to the one that matches them together (i.e. this matching also goes through the boundary). This symmetry may reduce BP's performance, hence we avoid it.

Regarding complexity, assuming we have $O(n)$ parallel resources with $O(1)$ space in each of them and hence $O(n)$ space in total, BP on the Tanner graph requires $O(T n)$ time to do $T$ iterations.
Furthermore, given a starting node $i \in \mv_{\text{D}}$, Dijkstra's algorithm computes its distance to all other nodes in $O(|\me_{\text{D}}| + |\mv_{\text{D}}| \log |\mv_{\text{D}}|)$. For the surface code, this requires $O(n \log n)$
time, assuming we can run $O(n)$ Dijkstra instances in parallel (spending $O(n)$ space in each Dijkstra and hence $O(n^2)$ space in total), i.e. one for each unsatisfied syndrome. This directly provides the weights associated to unsatisfied syndrome pairs that we need to initialize the BP approach to MWPM. In order to obtain the weights associated to matching an unsatisfied syndrome to the boundary, we need to minimize over all the  Dijkstra paths joining its associated node to a virtual node. 
This adds an extra $O(n)$ time, assuming we can do it for all unsatisfied syndromes in parallel.
Overall, adding correlations via this procedure requires $O(T n + n \log n)$
time.

\section{Results}
\label{sec:res}

The core of our results are two BP-based algorithms: BP4M, BP4MF. Additionally, we build also a two-stage decoder BP4M+M(atching) and one that incorporates correlations B-BP4MF.
Table \ref{table: complexities} lists the time complexity of each algorithm. In the naive implementation we have adopted for simplicity, each algorithm requires $O(n^2)$ space in the worst case. We discuss space complexity further in Section \ref{sec: conclusion}.


\vspace{-0.3cm}

\subsection{Belief Propagation for Matching (BP4M)}

Our first method is \textbf{belief propagation for matching (BP4M)}. Herein, we perform message-passing for a number of iterations $T$, performing marginalization at each iteration and, provided we converge in several iterations, we keep the error candidate with lowest weight among them. Once the iterations are over, and provided the last one did not converge, we perform forced convergence in order to get some error candidate for each syndrome.
Lastly, if forced convergence was used, we compare the weight of the lowest-weight candidate coming from marginalization with that of forced convergence and output the error candidate with smallest weight between them. Otherwise, we output the lowest-weight candidate coming from marginalization.

The complexity of each iteration of BP4M is that of message-passing itself $O(n)$ plus that of the marginalization $O(1)$ for a total of $O(nT)$ when finishing the iterations. After that, forced convergence requires $O(n \log n)$, and we get an overall complexity of $O(nT + n \log n)$. 

\subsection{BP4M Forced (BP4MF)}

The second method is called \textbf{belief propagation for matching forced (BP4MF)}.
In this approach, we perform message-passing for a number of iterations $T$, performing forced convergence at each iteration and keeping the error candidate with lowest weight among them. For the same number of iterations $T$, we expect this approach to perform better than BP4M, since the weight of the final output will be upper bounded by that of BP4M. This is the case since, if marginalization converges, then forced convergence will output the same error candidate.

The complexity of each iteration of BP4MF is that of message-passing itself $O(n)$ plus that of forced convergence $O(n \log n)$ for a total of $O(Tn \log (n))$ when finishing the iterations.
Before performing forced convergence, we can marginalize and check for convergence for $O(n)$, and avoid the $O(n \log n)$ cost of forced convergence. The number of times forced convergence cannot be avoided diminishes with the physical error rate (see Figure \ref{fig: calls matching}).

\subsection{BP4M + Matching (BP4M+M)}

The third method is \textbf{belief propagation for matching plus matching (BP4M+M)}. This is a two-stage decoder, where the first stage is running BP4M and the second one vanilla MWPM. In this approach, we reproduce BP4M while we are performing iterations and, once we have reached $T$ iterations, we check whether we have converged. If that is the case, then we output the converged error candidate with minimal weight like BP4M does. Otherwise, we perform MWPM. This method is somehow similar to~\cite{higgott2023improved}, although the soft information from BP4M is not used to reweigh the MWPM graph.

The complexity of each iteration of BP4M+M is that of BP4M, that is, $O(nT)$ after the iterations have finished. If we do not obtain an error candidate that matches the observed syndrome after $T$ iterations, then we use matching for an extra complexity of $O(n^3 \log n)$. As a result, if we call $r_{nc}$ the \textbf{ratio of non-convergence}, i.e. the quotient of the number of times we do not converge and the number of errors we have generated, then the expected runtime is $O((1-r_{nc})nT + r_{nc} n^3 \log n)$.
As we show in Fig.~\ref{fig: calls matching}, $r_{nc}$ diminishes with the physical error rate and we get, for $p=0.02$ and $d=3,5,7,9,11$ under the code capacity noise model, that $r_{nc} \leq 0.1$.

\subsection{Belief-BP4MF (B-BP4MF)}

The fourth method is \textbf{quaternary belief propagation plus belief propagation for matching forced} or \textbf{belief-BP4MF (B-BP4MF)}. This is a two-stage decoder, where the first stage runs quaternary BP. If the first stage converges, we take its hard decision as output. Otherwise, we run Dijkstra's algorithm (see Section \ref{sec: correlated}) in order for the BP4MF initial log-likelihoods to take into account correlations, and we then run BP4MF. One can substitute BP4MF by BP4M in order to get another decoder, although we only simulate the BP4MF case. This methods are the belief propagation for matching counterparts of belief-matching \cite{higgott2023improved}.



Since the time complexity of quaternary BP together with our Dijkstra subroutine is $O(T n + n\log n)$ and that of BP4MF is $O(T n\log n)$, the overall complexity of B-BP4MF is $O(T n \log n)$. We use the same number of iterations $T=75$ for both quaternary BP and BP4MF.


\subsection{Numerical Results}

We benchmark our decoding algorithms across code distances $d \in \{3, 5, 7, 9, 11\}$ using both code capacity and circuit-level noise models:

For code capacity noise, we evaluate the unrotated surface code subject to a standard depolarizing channel with physical error rates $p \in [0.02, 0.18]$.

For circuit-level noise, we consider both the unrotated  and rotated memory-Z surface codes. The  circuits are constructed using Stim's \cite{gidney2021stim} standard $\texttt{surface\_code:rotated\_memory\_z }$ and  $\texttt{surface\_code:unrotated\_memory\_z }$ generation, employing $R=d$ syndrome-extraction rounds, ending in a final data-qubit measurement. The noise model injects a depolarizing error (probability $p$) after every Clifford gate and onto data qubits before each syndrome extraction, with state preparation (reset) and measurement flip probabilities similarly set to $p$. We sweep the physical error rate from $p=0.002$ to $0.012$ for most configurations, while for the B+BP4MF decoder, we evaluate  code distances $d \in \{5,9,13\}$ across rates from $p=0.003$ to $0.012$.


We implemented our code with Numpy and C++.
Simulations were performed on a Linux workstation running Ubuntu 24.04.3 LTS (kernel 6.8.0-94-generic), equipped with an AMD EPYC 9274F (24 cores / 48 threads, single socket; up to 4.05 GHz) and 188 GB RAM, and 2× NVIDIA RTX 6000 Ada Generation GPUs (48 GB VRAM each). For comparison against vanilla MWPM, we utilize PyMatching~\cite{higgott2022pymatching}. This library also serves as the underlying MWPM subroutine within our BP4M+M decoder. We have also used the QECSIM library~\cite{qecsim}.


\begin{figure}[!t]
    \centering
    \resizebox{\columnwidth}{!}{
\begin{tikzpicture}

\definecolor{tabpurple}{HTML}{9467BD}
\definecolor{tabcyan}  {HTML}{17BECF}
\definecolor{taborange}{HTML}{FF7F0E}

\pgfdeclareplotmark{trianglefilled}{%
  \pgfpathmoveto{\pgfqpoint{0pt}{\pgfplotmarksize}}%
  \pgfpathlineto{\pgfqpoint{-\pgfplotmarksize}{-\pgfplotmarksize}}%
  \pgfpathlineto{\pgfqpoint{\pgfplotmarksize}{-\pgfplotmarksize}}%
  \pgfpathclose\pgfusepathqfillstroke%
}

\pgfplotsset{
  mBP4M/.style={color=tabpurple, line width=3.5pt,
    dash pattern=on 12.95pt off 5.6pt,
    mark=*, mark size=4.5pt,
    mark options={solid, fill=tabpurple, draw=tabpurple, line width=1.2pt}},
  mBP4MF/.style={color=tabcyan, line width=3.5pt,
    dash pattern=on 12.95pt off 5.6pt,
    mark=*, mark size=4.5pt,
    mark options={solid, fill=tabcyan, draw=tabcyan, line width=1.2pt}},
  mBP4MM/.style={color=taborange, line width=3.5pt,
    dash pattern=on 12.95pt off 5.6pt,
    mark=*, mark size=4.5pt,
    mark options={solid, fill=taborange, draw=taborange, line width=1.2pt}},
  mPyM/.style={color=black, line width=3.5pt,
    dash pattern=on 3.5pt off 5.775pt,
    mark=trianglefilled, mark size=4.5pt,
    mark options={solid, fill=black, draw=black, line width=1.2pt}},
  panel/.style={
    scale only axis=true,
    width=21.67cm, height=9.58cm,
    xmin=2.5, xmax=11.5,
    ymin=0.011229387237043803, ymax=102.89613064088982,
    xtick={3,5,7,9,11},
    scaled x ticks=false,
    ytick={0.1,1,10,100},
  minor ytick={0.02,0.03,0.04,0.05,0.06,0.07,0.08,0.09,0.2,0.3,0.4,0.5,
              0.6,0.7,0.8,0.9,2,3,4,5,6,7,8,9,20,30,40,50,60,70,80,90},
    label style={font=\fontsize{23}{27}\selectfont},
    title style={font=\fontsize{23}{27}\selectfont},
    tick label style={font=\fontsize{23}{27}\selectfont},
    tick align=outside,
    xtick pos=left, ytick pos=left,
    grid=both,
    grid style      ={mplgrid, opacity=0.7, dotted, line width=1pt},
    minor grid style={mplgrid, opacity=0.7, dotted, line width=1pt},
    axis on top=false,
  },
}

\begin{axis}[ymode=log, axis background/.style={fill=white},
  panel, name=axtop,
  title={p = 0.06},
  ylabel={run time (s)},
  xticklabel=\empty]

  \addplot[mBP4M, forget plot] table[x=d, y=t] {
      d    t
      3    0.028
      5    0.21
      7    0.758
      9    1.858
      11   3.692
  };

  \addplot[mBP4MF, forget plot] table[x=d, y=t] {
      d    t
      3    0.151372578
      5    0.453879055
      7    1.123324146
      9    2.764891584
      11   6.295028469
  };

  \addplot[mBP4MM, forget plot] table[x=d, y=t] {
      d    t
      3    0.697
      5    2.435
      7    4.856
      9    8.306
      11   13.16
  };

  \addplot[mPyM, forget plot] table[x=d, y=t] {
      d    t
      3    19.93898487
      5    23.10552997
      7    29.76171121
      9    43.02934156
      11   67.96826448
  };

\end{axis}

\begin{axis}[ymode=log, axis background/.style={fill=white},
  panel, name=axbot,
  at={(axtop.south west)}, anchor=north west, yshift=-1.39cm,
  title={p = 0.03},
  xlabel={code distance},
  ylabel={run time (s)},
  legend style={
    at={(0.43685,-0.2763)}, anchor=north,
    font=\fontsize{16}{18}\selectfont,
    fill=white, fill opacity=0.8, text opacity=1,
    draw=black!25, rounded corners=2pt,
    column sep=10pt,
  },
  legend columns=4,
  legend cell align=left,
  legend image code/.code={%
    \draw[mark repeat=2, mark phase=2, #1]
      plot coordinates {(0cm,0cm) (0.65cm,0cm) (1.3cm,0cm)};%
  },
]

  \addplot[mBP4M, forget plot] table[x=d, y=t] {
      d    t
      3    0.017
      5    0.085
      7    0.366
      9    1.013
      11   1.858
  };

  \addplot[mBP4MF, forget plot] table[x=d, y=t] {
      d    t
      3    0.088832961
      5    0.261452984
      7    0.622998373
      9    1.106465926
      11   2.105452149
  };

  \addplot[mBP4MM, forget plot] table[x=d, y=t] {
      d    t
      3    0.205
      5    0.65
      7    1.271
      9    2.065
      11   3.29
  };

  \addplot[mPyM, forget plot] table[x=d, y=t] {
      d    t
      3    19.63884134
      5    22.12830322
      7    27.82044971
      9    40.26510677
      11   63.86552271
  };

  \addlegendimage{mBP4M}   \addlegendentry{BP4M}
  \addlegendimage{mBP4MF}  \addlegendentry{BP4MF}
  \addlegendimage{mBP4MM}  \addlegendentry{BP4M+M}
  \addlegendimage{mPyM}    \addlegendentry{PyMatching}

\end{axis}
\end{tikzpicture}}
    \caption{Code capacity average runtimes (over 1 million errors) compared to MWPM. We use $T=25$ iterations for our decoders.}
    \label{fig: runtime}
\end{figure}

\begin{figure*}[t]
\centering
\begin{subfigure}{0.49\textwidth}\centering
  \textbf{}\\[0.5em]
  \resizebox{\linewidth}{!}{\begin{tikzpicture}

\newcommand{\cpms}{4.5pt}
\definecolor{mplpink}{HTML}{008080}

\pgfplotsset{
  cpcurve/.style={
    color=#1,
    line width=3.5pt,
    mark=*,
    mark size=\cpms,
    mark options={solid, fill=#1, draw=#1, line width=1.2pt},
    error bars/y dir=both,
    error bars/y explicit,
    error bars/error bar style={color=#1, line width=2.5pt},
    error bars/error mark=-,
    error bars/error mark options={color=#1, mark size=3pt, line width=2.5pt},
  },
  cpkey/.style={
    color=#1, line width=3.5pt, mark=*, mark size=\cpms,
    mark options={solid, fill=#1, draw=#1, line width=1.2pt},
  },
}

\begin{axis}[
  xmode=log, ymode=log,
  axis background/.style={fill=white},
  width=22.86cm, height=15.9cm,
  scale only axis=false,
  xmin=0.028, xmax=0.21,
  ymin=2e-06, ymax=0.60,
  xlabel={physical error rate},
  ylabel={LER},
  xlabel style={font=\fontsize{25}{30}\selectfont},
  ylabel style={font=\fontsize{30}{34}\selectfont},
  tick label style={font=\fontsize{20}{24}\selectfont},
  xtick={0.03,0.06,0.1,0.14,0.18},
  xticklabels={$0.03$,$0.06$,$0.1$,$0.14$,$0.18$},
  minor xtick={0.025,0.035,0.05,0.07,0.08,0.09,0.11,0.12,0.13,0.14,0.16,0.17,0.18},
  ytick={1e-5,1e-4,1e-3,1e-2,1e-1},
  minor ytick={3e-06,4e-06,5e-06,6e-06,7e-06,8e-06,9e-06,
               2e-05,3e-05,4e-05,5e-05,6e-05,7e-05,8e-05,9e-05,
               0.0002,0.0003,0.0004,0.0005,0.0006,0.0007,0.0008,0.0009,
               0.002,0.003,0.004,0.005,0.006,0.007,0.008,0.009,
               0.02,0.03,0.04,0.05,0.06,0.07,0.08,0.09,0.2,0.3,0.4,0.5},
  tick align=outside,
  xtick pos=left, ytick pos=left,
  grid=both,
  grid style      ={mplgrid, dotted, line width=1pt},
  minor grid style={mplgrid, dotted, line width=1pt},
  axis on top=false,
  legend style={
    at={(0.98,0.02)}, anchor=south east,
    font=\fontsize{18}{22}\selectfont,
    fill=white, fill opacity=0.9, text opacity=1,
    draw=black!25, rounded corners=2pt,
    row sep=1pt, column sep=8pt,
  },
  legend columns=2,
  legend cell align=left,
  legend image code/.code={%
    \draw[mark repeat=2, mark phase=2, #1]
      plot coordinates {(0cm,0cm) (0.55cm,0cm) (1.1cm,0cm)};%
  },
]

  \addplot[forget plot, black,    mpldotted, line width=3.5pt]
    coordinates {(0.155,1e-8) (0.155,1)};   
  \addplot[forget plot, mplbrown, mpldotted, line width=3.5pt]
    coordinates {(0.15,1e-8) (0.15,1)};   

  \addplot[cpcurve=mplgreen, forget plot]
    table[x=p, y=ler, y error plus=eplus, y error minus=eminus] {
      p        ler              eplus            eminus
     
      0.03     0.001189         0.000087         0.000083
      0.04     0.003082         0.000225         0.000214
      0.05     0.006250         0.000456         0.000433
      0.06     0.011660         0.000849         0.000806
      0.07     0.01856         0.000461         0.000391
      0.08     0.030544         0.002200         0.002094
      0.09     0.041415         0.002965         0.002825
      0.10     0.057524         0.004082         0.003893
      0.11     0.078536         0.005506         0.005260
      0.12     0.102020         0.007054         0.006752
      0.13     0.131596         0.008938         0.008577
      0.14     0.151814         0.010182         0.009788
      0.15     0.189466         0.012401         0.011964
      0.16     0.216685         0.013926         0.013470
      0.17     0.264340         0.016429         0.015964
      0.18     0.283366         0.017367         0.016909
    };

  \addplot[cpcurve=mplpurple, forget plot]
    table[x=p, y=ler, y error plus=eplus, y error minus=eminus] {
      p        ler              eplus            eminus
      
      0.03     0.000047         0.000005         0.000005
      0.04     0.000225         0.000016         0.000016
      0.05     0.000806         0.000059         0.000056
      0.06     0.002206         0.000162         0.000153
      0.07     0.004773         0.000349         0.000331
      0.08     0.010105         0.000737         0.000699
      0.09     0.018102         0.001313         0.001249
      0.10     0.031541         0.002271         0.002161
      0.11     0.048541         0.003462         0.003299
      0.12     0.073535         0.005170         0.004937
      0.13     0.101420         0.007015         0.006715
      0.14     0.146413         0.009852         0.009468
      0.15     0.188715         0.012358         0.011922
      0.16     0.240674         0.015213         0.014749
      0.17     0.296384         0.017987         0.017538
      0.18     0.330469         0.019531         0.019113
    };

  \addplot[cpcurve=mplpink, forget plot]
    table[x=p, y=ler, y error plus=eplus, y error minus=eminus] {
      p        ler              eplus            eminus
      0.03     0.000009         0.000001         0.000001
      0.04     0.000057         0.000006         0.000005
      0.05     0.000254         0.000018         0.000018
      0.06     0.000854         0.000062         0.000060
      0.07     0.002182         0.000159         0.000152
      0.08     0.005156         0.000376         0.000358
      0.09     0.010317         0.000752         0.000714
      0.10     0.019138         0.0001387        0.001319
      0.11     0.035476         0.002549         0.002426
      0.12     0.059256         0.004200         0.004007
      0.13     0.089606         0.006241         0.005968
      0.14     0.131944         0.008959         0.008598
      0.15     0.188218         0.012330         0.011894
      0.16     0.246063         0.015495         0.015030
      0.17     0.307692         0.018513         0.018072
      0.18     0.386250         0.021792         0.021465
    };
  \addplot[forget plot, mplbrown, mpldotted, line width=3.5pt]
    coordinates {(0.00,2.9341616804130624e-05) (0.0057,0.5462227573011053)};
  \addplot[forget plot, black, line width=3.5pt,
           dash pattern=on 12.95pt off 5.6pt]
    coordinates {(0.0072,2.9341616804130624e-05) (0.0072,0.5462227573011053)};

  \addplot[forget plot, mplbrown, mpldashdot, line width=3.5pt,
           domain=0.018:0.21, samples=2] {x};

  \node[anchor=north west,
        draw=black!25, fill=white, fill opacity=0.9, text opacity=1,
        rounded corners=2pt, inner sep=6pt,
        font=\fontsize{18}{22}\selectfont]
    at (rel axis cs:0.02,0.98) {Code Capacity};
    \node[anchor=south west, black, font=\bfseries\fontsize{20}{24}\selectfont]
    at (axis cs:0.156, 9e-5) {MWPM};
  \node[anchor=south west, black, font=\bfseries\fontsize{20}{24}\selectfont]
    at (axis cs:0.157, 2e-4) {0.155};
  \node[anchor=south west, mplbrown, font=\bfseries\fontsize{20}{24}\selectfont]
    at (axis cs:0.1, 9e-5) {B-BP4MF};
  \node[anchor=south west, mplbrown, font=\bfseries\fontsize{20}{24}\selectfont]
    at (axis cs:0.115, 2e-4) {0.15};

\end{axis}
\end{tikzpicture}}
  \caption{}
  \label{fig: b+bp4mf_code_capacity}
\end{subfigure}\hfill
\begin{subfigure}{0.49\textwidth}\centering
  \textbf{}\\[0.5em]
  \resizebox{\linewidth}{!}{
\begin{tikzpicture}
\definecolor{mplpink}{HTML}{008080}

%
\newcommand{\cpms}{4.5pt}

\pgfplotsset{
  cpcurve/.style={
    color=#1,
    line width=3.5pt,
    mark=*,
    mark size=\cpms,
    mark options={solid, fill=#1, draw=#1, line width=1.2pt},
    error bars/y dir=both,
    error bars/y explicit,
    error bars/error bar style={color=#1, line width=2.5pt},
    error bars/error mark=-,
    error bars/error mark options={color=#1, mark size=3pt, line width=2.5pt},
  },
  cpkey/.style={
    color=#1, line width=3.5pt, mark=*, mark size=\cpms,
    mark options={solid, fill=#1, draw=#1, line width=1.2pt},
  },
}

\begin{axis}[
  xmode=log, ymode=log,
  axis background/.style={fill=white},
  width=22.86cm, height=15.24cm,          
  scale only axis=false,
  xmin=0.0018286156535236553, xmax=0.01312468257271731,
  ymin=8.71099747483038e-06, ymax=0.7212928480668759,
  xlabel={physical error rate },
  ylabel={},
  label style={font=\fontsize{21}{25}\selectfont},
  tick label style={font=\fontsize{20}{24}\selectfont},
  xtick={0.002,0.003,0.004,0.006,0.01},
  xticklabels={$2\times10^{-3}$,$3\times10^{-3}$,$4\times10^{-3}$,
               $6\times10^{-3}$,$10^{-2}$},
  minor xtick={0.005,0.007,0.008,0.009},
  ytick={1e-5,1e-4,1e-3,1e-2,1e-1},
  minor ytick={9e-06,2e-05,3e-05,4e-05,5e-05,6e-05,7e-05,8e-05,9e-05,
              0.0002,0.0003,0.0004,0.0005,0.0006,0.0007,0.0008,0.0009,
              0.002,0.003,0.004,0.005,0.006,0.007,0.008,0.009,0.02,0.03,
              0.04,0.05,0.06,0.07,0.08,0.09,0.2,0.3,0.4,0.5,0.6,0.7},
  tick align=outside,
  xtick pos=left, ytick pos=left,
  grid=both,
  grid style      ={mplgrid, dotted, line width=1pt},
  minor grid style={mplgrid, dotted, line width=1pt},
  axis on top=false,
  legend style={
    at={(0.98,0.02)}, anchor=south east,
    font=\fontsize{18}{22}\selectfont,
    fill=white, fill opacity=0.9, text opacity=1,
    draw=black!25, rounded corners=2pt,
    row sep=1pt, column sep=8pt,
  },
  legend columns=2,
  legend cell align=left,
  legend image code/.code={%
    \draw[mark repeat=2, mark phase=2, #1]
      plot coordinates {(0cm,0cm) (0.55cm,0cm) (1.1cm,0cm)};%
  },
]

  \addplot[forget plot, mplbrown, mpldotted, line width=3.5pt]
    coordinates {(0.0061,8.71099747483038e-06) (0.0061,0.7212928480668759)};
  \addplot[forget plot, black,    mpldotted, line width=3.5pt]
    coordinates {(0.0077,8.71099747483038e-06) (0.0077,0.7212928480668759)};

  \addplot[cpcurve=mplgreen, forget plot]
    table[x=p, y=ler, y error plus=eplus, y error minus=eminus] {
      p        ler              eplus            eminus
      0.002    0.00114729021    8.364480499e-05  7.942312701e-05
      0.003    0.003559951241   0.0002565912279  0.0002438097495
      0.004    0.008690585609   0.0006151603632  0.0005851881577
      0.005    0.01633265399    0.001101463164   0.001050594438
      0.006    0.02766170058    0.001803198537   0.001723499895
      0.007    0.04083806818    0.00247862927    0.002377725012
      0.008    0.05878155048    0.003311523189   0.00318851148
      0.009    0.08189174107    0.004286535183   0.004142265505
      0.01     0.1018337674     0.005100167945   0.004939908046
      0.011    0.1340846121     0.003991377014   0.003919251808
      0.012    0.1597826087     0.004699615859   0.004619137854
    };

  \addplot[cpcurve=mplpurple, forget plot]
    table[x=p, y=ler, y error plus=eplus, y error minus=eminus] {
      p        ler              eplus            eminus
      0.002    0.0001158574564  8.486274336e-06  8.055640382e-06
      0.003    0.000865275784   6.325591052e-05  6.005420654e-05
      0.004    0.003457828444   0.0002476330122  0.0002353714212
      0.005    0.01061649134    0.0007368340089  0.0007016642189
      0.006    0.02509973404    0.001645450265   0.00157213696
      0.007    0.05132378472    0.003052965831   0.002932464469
      0.008    0.08500976562    0.004467787719   0.004317426532
      0.009    0.1290893555     0.005985664908   0.005817768991
      0.01     0.1841517857     0.00737204354    0.007208735802
      0.011    0.2466408269     0.01116633222    0.01092337309
      0.012    0.2984496124     0.0124650544     0.01225031935
    };

  \addplot[cpcurve=mplpink, forget plot]
    table[x=p, y=ler, y error plus=eplus, y error minus=eminus] {
      p        ler              eplus            eminus
      0.002    1.739621458e-05  3.197503506e-06  2.820987478e-06
      0.003    0.0002997680757  2.19211734e-05   2.081071308e-05
      0.004    0.001927160266   0.0001408770647  0.0001337541762
      0.005    0.008127259814   0.0005906070571  0.0005610896286
      0.006    0.02569110577    0.001831422381   0.001743115756
      0.007    0.06117876838    0.004194023862   0.004006606237
      0.008    0.1229383681     0.007872601497   0.007568432078
      0.009    0.1966145833     0.01162780338    0.01126069284
      0.01     0.2769325658     0.01465064694    0.01430972402
      0.011    0.3593911249     0.01753195468    0.01726218832
      0.012    0.4119883041     0.01909795842    0.01890652831
    };

  \addplot[forget plot, mplbrown, mpldashdot, line width=3.5pt,
           domain=0.0018286156535236553:0.01312468257271731, samples=2] {x};

  \node[anchor=north west,
        draw=black!25, fill=white, fill opacity=0.9, text opacity=1,
        rounded corners=2pt, inner sep=6pt,
        font=\fontsize{18}{22}\selectfont]
    at (rel axis cs:0.02,0.98) {Unrotated Surface memory-Z};

  \node[anchor=south east, mplbrown,
        font=\bfseries\fontsize{20}{24}\selectfont]
    at (axis cs:0.00600, 1.876e-4) {B-BP4MF};
  \node[anchor=south east, mplbrown,
        font=\bfseries\fontsize{20}{24}\selectfont]
    at (axis cs:0.00600, 4.247e-4) {0.0061};
  \node[anchor=south west, black,
        font=\bfseries\fontsize{20}{24}\selectfont]
    at (axis cs:0.00790, 1.876e-4) {MWPM};
  \node[anchor=south west, black,
        font=\bfseries\fontsize{20}{24}\selectfont]
    at (axis cs:0.00790, 4.247e-4) {0.0077};

  \addlegendimage{cpkey=mplgreen}   \addlegendentry{$d = 5$}
  \addlegendimage{cpkey=mplpink}  \addlegendentry{$d = 13$}
  \addlegendimage{cpkey=mplpurple}  \addlegendentry{$d = 9$}
  \addlegendimage{mplbrown, mpldashdot, line width=3.5pt}
                                    \addlegendentry{LER = p}

\end{axis}
\end{tikzpicture}}
  \caption{}
  \label{fig: b+bp4mf_unrotated_surface}
\end{subfigure}
\caption{B-BP4MF threshold results:  Code capacity unrotated surface code with 25+25 iterations and circuit-level noise unrotated memory-Z surface code with 75+75 iterations. The horizontal red dotted line is the pseudo-threshold and the vertical dotted line indicates the achieved threshold.} 
\label{fig: b+bp4mf}
\end{figure*}

\begin{figure*}[t]
\centering
\begin{minipage}{0.49\textwidth}\centering
  \textbf{}\\[0.5em]
  \resizebox{\linewidth}{!}{\input{plots/plot_tex/threshold_mbp4f_rotated}}
 
\end{minipage}\hfill
\begin{minipage}{0.49\textwidth}\centering
  \textbf{}\\[0.5em]
  \resizebox{\linewidth}{!}{\input{plots/plot_tex/threshdol_unrotated_mbp4mf}}
  
\end{minipage}
\caption{Memory-BP4MF ($\alpha=0.85$) circuit-level noise (rotated and unrotated) memory-Z surface code threshold plots with 50 iterations.
  The horizontal red dotted line is the pseudo-threshold and the vertical dotted line indicates the achieved threshold.} 
\label{fig:  mbp4mf}
\end{figure*}

Since all our algorithms perform message-passing, their complexity and performance depend on the number of iterations $T$ they use.
We consider the following number of iterations: $T=25$ for all code capacity simulations except for B-BP4MF, where we do
$T=25$ for quaternary BP and $T=25$ for BP4MF. $T=50$ for all circuit-level noise simulations except for B-BP4MF, where we do
$T=75$ for quaternary BP and $T=75$ for BP4MF. Table \ref{table: complexities} summarizes the number of iterations.

\subsubsection{Comparison to MWPM}

 All of our algorithms have thresholds comparable to that of MWPM both for code capacity (Fig.~\ref{fig: threshold plots}, \ref{fig: threshold bp4m+m} and \ref{fig: b+bp4mf_code_capacity}) and for circuit-level noise (e.g. Fig.~\ref{fig: threshold circuit level noise}). See Table \ref{table: complexities} for a summary of threshold across all algorithms and noise models. Furthermore, these comparable thresholds are achieved in a faster runtime than MWPM (Fig.~\ref{fig: runtime}).

 All of our algorithms show good performance. One can see this in Fig.~\ref{fig: performance comparison} for $d=9,11$. As expected, for the same number of iterations T, we get that BP4MF performs better than BP4M. When considering different values of $T$, we have noticed that, even in the intermediate iterations where we do not converge, forced convergence typically provides meaningful error candidates. In particular, we have concluded from our experiments that exploring more forced convergence candidates is better in terms of performance than running more iterations and only exploring the converged error candidates. Another conclusion that we draw from these experiments is that a low number of iterations already provide good results, with the performance improving slowly as $T$ grows. Moreover, these improvements become more prominent as the physical error rate diminishes. Our experiments with $T$ were performed under code capacity noise.

 When we converge via marginalization, the performance is comparable to that of MWPM. 
    The performance of BP4M+M (Fig.~\ref{fig: threshold bp4m+m}) is almost indistinguishable from that of MWPM
    (Fig.~\ref{fig: performance comparison}). This indicates that, when our algorithms converge via marginalization, their performance is comparable to that of MWPM. To support this, we include in Fig.~\ref{fig: pseudo ler} the \textbf{converged logical error rate} or \textbf{converged LER}, which consists of the logical errors ignoring the cases where we do not converge. Moreover, the number of instances where we do not converge via marginalization diminishes with the physical error rate (Fig.~\ref{fig: calls matching}).

\subsubsection{BP Improvements}

Known improvements for BP on the Tanner graph can also be advantageous for our algorithms. 
An example of this is adding memory \cite{kuo2022exploiting}. More specifically, when updating the messages from variable nodes to factor nodes, we add a tunable parameter $\alpha$, getting the following equation
$$ 
\mvc^{(t+1)} = \priorlog_{\vnode} + \frac{1}{\alpha} \sum_{\cnode' \in \nn(\vnode) \smin \cnode} \mkern-20mu m_{\cnode' \to \vnode}^{(t)}. 
$$
By doing a grid search with $\alpha = 0.6 + k \cdot 0.05$ and $k \in \{0,1,\dots,20\}$, we manage to improve the performance of BP4MF (see Figure \ref{fig: mbp4mf}). We refer to this algorithm as Memory-BP4MF.

We have tried a couple more approaches that did not yield significant gains, at least in the code capacity setup. This includes serial scheduling of the messages and adding memory during marginalization.

 \section{Conclusion}
 \label{sec: conclusion}

We have shown that the absence of a threshold for standard belief
propagation (BP) on the surface code is not intrinsic to the algorithm,
but a consequence of the graph on which messages are passed. Moving the
message-passing to the decoding graph, the graph on which minimum weight
perfect matching (MWPM) operates, recovers threshold behavior, despite the
short cycles present in the latter graph.

Our algorithms achieve thresholds of $0.117$ (BP4M) and $0.134$ (BP4MF)
under code capacity noise, against $0.155$ for MWPM, and of $0.0050$ and
$0.0055$ under circuit-level noise, against $0.0072$ and $0.0077$ for the
rotated and unrotated memory-Z experiments. The forced convergence
strategy of BP4MF, which guarantees a valid error candidate at every
iteration, is what closes most of the gap at higher distances. Adding
correlations in the style of belief-matching closes it further, raising
the code capacity threshold to $0.150$, at the cost of a second BP stage.
Used instead as a pre-decoder that defers to matching only when
marginalization fails, BP4M+M attains a threshold of $0.157$ while calling
MWPM on a small and rapidly decreasing fraction of syndromes, and is
faster than PyMatching across the distances we considered. We also find
that a small number of iterations suffices, and that heuristics developed
for BP on the Tanner graph transfer to our setting: adding memory improves
the circuit-level threshold of BP4MF to $0.0059$.

A gap to MWPM remains. We attribute it primarily to the greedy nature of
forced convergence, which fixes variables serially by reliability and
therefore does not recover the globally optimal matching that Blossom
provides, and to the independence assumption underlying the edge priors. Narrowing this gap without giving up the low per-iteration
cost is, in our view, the main open problem raised by this work.

Several directions follow. The X and Z matchings are currently decoded
independently, and improving their interaction is the natural route towards
correlated decoding. The matching-based approaches to non-graphlike codes
\cite{bombin2006topological, kubica2023efficient, brown2023conservation} may
allow our algorithms to be applied beyond graphlike codes. Further BP
heuristics are worth exploring, in particular concatenating several BP
instances in series or using ensembles, as Relay BP does
\cite{muller2025improved}. Our space complexity is $O(n^2)$ because we
build the complete decoding graph. Sparse Blossom \cite{higgott2025sparse}
shows that this can be avoided for matching, and doing so in our setting
would be practically relevant. Finally, the per-iteration cost of our
algorithms is $O(n)$ with $O(n)$ parallel resources and the message
updates are local, which makes an FPGA or ASIC implementation a natural
next step.

\section*{Acknowledgements}

We thank  Francisco García Herrero for feedback on the manuscript and Henry D. Pfister for useful discussions.
This research was supported by the Munich Quantum Valley, which is supported by the Bavarian state government with funds from the Hightech Agenda Bayern Plus.
Luca Menti's research was funded by a scholarship awarded by the German Academic Exchange Service (DAAD).

We acknowledge the EuroHPC Joint Undertaking for awarding this project access to the EuroHPC supercomputer LUMI, hosted by CSC (Finland) and the LUMI consortium through a EuroHPC Regular Access call.

\bibliography{__main}

\newpage
\section{Appendix}
\label{sec: appendix}
\captionsetup{type=figure} 
\vspace{4pt}

\centering
\begin{subfigure}[b]{0.32\textwidth}
\centering
\resizebox{\linewidth}{!}
{

\begin{tikzpicture}

\pgfplotsset{
  sqcurve/.style={
    color=#1,
    line width=3.5pt,
    mark=square*,
    mark size=4.5pt,
    mark options={fill=#1, draw=#1, line width=1.2pt},
  },
  sqkey/.style={
    color=#1, line width=3.5pt, mark=square*, mark size=4.5pt,
    mark options={fill=#1, draw=#1, line width=1.2pt},
  },
}

\begin{axis}[
  width=15cm, height=13.75cm,
  scale only axis=false,
  axis background/.style={fill=white},
  ymode=log,
  xmin=0.035, xmax=0.065,
  ymin=1.7419849063869548e-06, ymax=0.064507007642831168,
  xlabel={physical error rate},
  ylabel={converged LER},
  xlabel style={font=\fontsize{21}{25}\selectfont},
  ylabel style={font=\fontsize{25}{29}\selectfont},
  tick label style={font=\fontsize{20}{24}\selectfont},
  xtick={0.04,0.05,0.06},
  scaled x ticks=false,
  xticklabel style={/pgf/number format/fixed, /pgf/number format/fixed zerofill,
                    /pgf/number format/precision=2},
  ytick={1e-5,1e-4,1e-3,1e-2},
  minor ytick={2e-06,3e-06,4e-06,5e-06,6e-06,7e-06,8e-06,9e-06,2e-05,
              3e-05,4e-05,5e-05,6e-05,7e-05,8e-05,9e-05,0.0002,0.0003,
              0.0004,0.0005,0.0006,0.0007,0.0008,0.0009,0.002,0.003,
              0.004,0.005,0.006,0.007,0.008,0.009,0.02,0.03,0.04,0.05,
              0.06},
  tick align=outside,
  xtick pos=left, ytick pos=left,
  grid=both,
  grid style      ={mplgrid, opacity=0.7, dotted, line width=1pt},
  minor grid style={mplgrid, opacity=0.7, dotted, line width=1pt},
  axis on top=false,
  legend style={
    at={(0.47,0.02)}, anchor=south,
    font=\fontsize{16}{20}\selectfont,
    fill=white, fill opacity=0.9, text opacity=1,
    draw=black!25, rounded corners=2pt,
    row sep=1pt, column sep=8pt,
  },
  legend columns=2,
  legend cell align=left,
  legend image code/.code={%
    \draw[mark repeat=2, mark phase=2, #1]
      plot coordinates {(0cm,0cm) (0.55cm,0cm) (1.1cm,0cm)};%
  },
]

  \addplot[sqcurve=mplblue, forget plot] table[x=p, y=ler] {
      p            ler
      0.06         0.039989407
      0.05         0.026427292
      0.04         0.017465427
  };

  \addplot[sqcurve=mplgreen, forget plot] table[x=p, y=ler] {
      p            ler
      0.06         0.012495182
      0.05         0.006679625
      0.04         0.004070815
  };

  \addplot[sqcurve=mplred, forget plot] table[x=p, y=ler] {
      p            ler
      0.06         0.002594431
      0.05         0.00127984
      0.04         0.000497321
  };

  \addplot[sqcurve=mplpurple, forget plot] table[x=p, y=ler] {
      p            ler
      0.06         0.000375532
      0.05         0.000162144
      0.04         5.36e-05
  };

  \addplot[sqcurve=mplorange, forget plot] table[x=p, y=ler] {
      p            ler
      0.06         4.04e-05
      0.05         2.08e-05
      0.04         2.81e-06
  };

  %

\end{axis}
\end{tikzpicture}}
\caption{}\label{fig: pseudo ler}
\end{subfigure}\hfill
\begin{subfigure}[b]{0.32\textwidth}
\centering
\resizebox{\linewidth}{!}
{
\begin{tikzpicture}

\pgfplotsset{
  scurve/.style={
    color=#1,
    line width=3.2pt,
    mark=square*,
    mark size=4.25pt,
    mark options={fill=#1, draw=#1, line width=1.1pt},
  },
  skey/.style={
    color=#1, line width=3.2pt, mark=square*, mark size=4.25pt,
    mark options={fill=#1, draw=#1, line width=1.1pt},
  },
}

\begin{axis}[
  width=15cm, height=12.75cm,
  scale only axis=false,
  axis background/.style={fill=white},
  ymode=log,
  xmin=0.015, xmax=0.19,
  ymin=0.005515877742573562, ymax=1.2809794433013166,
  xlabel={physical error rate},
  ylabel={fraction of calls to MWPM},
  label style={font=\fontsize{20}{24}\selectfont},
  tick label style={font=\fontsize{20}{24}\selectfont},
  xtick={0.02,0.06,0.10,0.14,0.18},
  scaled x ticks=false,
  xticklabel style={/pgf/number format/fixed, /pgf/number format/fixed zerofill,
                    /pgf/number format/precision=2},
  ytick={1e-2,1e-1,1},
  minor ytick={0.006,0.007,0.008,0.009,
               0.02,0.03,0.04,0.05,0.06,0.07,0.08,0.09,
               0.2,0.3,0.4,0.5,0.6,0.7,0.8,0.9},
  tick align=outside,
  xtick pos=left, ytick pos=left,
  grid=both,
  grid style      ={mplgrid, opacity=0.7, dotted, line width=1pt},
  minor grid style={mplgrid, opacity=0.7, dotted, line width=1pt},
  axis on top=false,
  legend style={
    at={(0.98,0.024)}, anchor=south east,
    font=\fontsize{17}{21}\selectfont,
    fill=white, fill opacity=0.9, text opacity=1,
    draw=black!25, rounded corners=2pt,
    row sep=1pt, column sep=8pt,
  },
  legend columns=2,
  legend cell align=left,
  legend image code/.code={%
    \draw[mark repeat=2, mark phase=2, #1]
      plot coordinates {(0cm,0cm) (0.55cm,0cm) (1.1cm,0cm)};%
  },
]

  \addplot[scurve=mplblue, forget plot] table[x=p, y=frac] {
      p            frac
      0.18         0.28272935
      0.17         0.277157045
      0.16         0.239577384
      0.15         0.224876442
      0.14         0.201012292
      0.13         0.178076063
      0.12         0.163827751
      0.11         0.128692962
      0.1          0.125133364
      0.09         0.103318798
      0.08         0.084207515
      0.07         0.072727273
      0.06         0.054463503
      0.05         0.039924512
      0.04         0.02784458
      0.03         0.015715453
      0.02         0.007065726
  };

  \addplot[scurve=mplgreen, forget plot] table[x=p, y=frac] {
      p            frac
      0.18         0.793412102
      0.17         0.739932886
      0.16         0.689427972
      0.15         0.644130956
      0.14         0.560522239
      0.13         0.505863956
      0.12         0.441502854
      0.11         0.390184711
      0.1          0.321577429
      0.09         0.264563512
      0.08         0.215290778
      0.07         0.175197029
      0.06         0.128272618
      0.05         0.087022457
      0.04         0.055179744
      0.03         0.031882288
      0.02         0.014060476
  };

  \addplot[scurve=mplred, forget plot] table[x=p, y=frac] {
      p            frac
      0.18         0.982029203
      0.17         0.954642098
      0.16         0.938346298
      0.15         0.901536137
      0.14         0.859779381
      0.13         0.803646308
      0.12         0.750372356
      0.11         0.669484702
      0.1          0.600569419
      0.09         0.51427967
      0.08         0.422598542
      0.07         0.337957848
      0.06         0.258031732
      0.05         0.182160129
      0.04         0.118786081
      0.03         0.067186
      0.02         0.029599
  };

  \addplot[scurve=mplpurple, forget plot] table[x=p, y=frac] {
      p            frac
      0.18         0.998007968
      0.17         0.997067449
      0.16         0.99253954
      0.15         0.980142566
      0.14         0.96850225
      0.13         0.940426194
      0.12         0.914609661
      0.11         0.862368766
      0.1          0.799773756
      0.09         0.720961855
      0.08         0.626832063
      0.07         0.519397298
      0.06         0.407900588
      0.05         0.299333304
      0.04         0.198198
      0.03         0.113743
      0.02         0.051049
  };

  \addplot[scurve=mplorange, forget plot] table[x=p, y=frac] {
      p            frac
      0.18         1
      0.17         0.999616564
      0.16         0.999377722
      0.15         0.997964894
      0.14         0.996051889
      0.13         0.991107406
      0.12         0.97895565
      0.11         0.956045356
      0.1          0.92081231
      0.09         0.866492746
      0.08         0.785152631
      0.07         0.682226845
      0.06         0.557040814
      0.05         0.422523
      0.04         0.288094
      0.03         0.16931
      0.02         0.0773107
  };

  \addlegendimage{skey=mplblue}    \addlegendentry{$d=3$}
  \addlegendimage{skey=mplpurple}  \addlegendentry{$d=9$}
  \addlegendimage{skey=mplgreen}   \addlegendentry{$d=5$}
  \addlegendimage{skey=mplorange}  \addlegendentry{$d=11$}
  \addlegendimage{skey=mplred}     \addlegendentry{$d=7$}

\end{axis}
\end{tikzpicture}}
\caption{}\label{fig: calls matching}
\end{subfigure}\hfill
\begin{subfigure}[b]{0.32\textwidth}
\centering
\resizebox{\linewidth}{!}
{\input{plots/plot_tex/threshold_bp_plus_matching_25_iter_fig}}
\caption{}\label{fig: threshold bp4m+m}
\end{subfigure}

\caption{Code capacity convergence performance and MWPM postprocessing:
(a) Converged LER. (b) Convergence fraction. (c) Performance of BP4M+M
with 25 iterations.}
\label{fig: convergence panels}

\vspace{7pt} 
\vspace{3 cm}
\begin{minipage}[t]{\columnwidth}
\centering
\resizebox{\linewidth}{!}{
\begin{tikzpicture}

\definecolor{tabpurple}{HTML}{9467BD}
\definecolor{tabcyan}  {HTML}{17BECF}
\definecolor{taborange}{HTML}{FF7F0E}

\pgfdeclareplotmark{plusfilled}{%
  \pgfpathmoveto{\pgfqpoint{-0.33333\pgfplotmarksize}{-\pgfplotmarksize}}%
  \pgfpathlineto{\pgfqpoint{0.33333\pgfplotmarksize}{-\pgfplotmarksize}}%
  \pgfpathlineto{\pgfqpoint{0.33333\pgfplotmarksize}{-0.33333\pgfplotmarksize}}%
  \pgfpathlineto{\pgfqpoint{\pgfplotmarksize}{-0.33333\pgfplotmarksize}}%
  \pgfpathlineto{\pgfqpoint{\pgfplotmarksize}{0.33333\pgfplotmarksize}}%
  \pgfpathlineto{\pgfqpoint{0.33333\pgfplotmarksize}{0.33333\pgfplotmarksize}}%
  \pgfpathlineto{\pgfqpoint{0.33333\pgfplotmarksize}{\pgfplotmarksize}}%
  \pgfpathlineto{\pgfqpoint{-0.33333\pgfplotmarksize}{\pgfplotmarksize}}%
  \pgfpathlineto{\pgfqpoint{-0.33333\pgfplotmarksize}{0.33333\pgfplotmarksize}}%
  \pgfpathlineto{\pgfqpoint{-\pgfplotmarksize}{0.33333\pgfplotmarksize}}%
  \pgfpathlineto{\pgfqpoint{-\pgfplotmarksize}{-0.33333\pgfplotmarksize}}%
  \pgfpathlineto{\pgfqpoint{-0.33333\pgfplotmarksize}{-0.33333\pgfplotmarksize}}%
  \pgfpathclose\pgfusepathqfillstroke%
}
\pgfdeclareplotmark{diamondfilled}{%
  \pgfpathmoveto{\pgfqpoint{0pt}{-\pgfplotmarksize}}%
  \pgfpathlineto{\pgfqpoint{\pgfplotmarksize}{0pt}}%
  \pgfpathlineto{\pgfqpoint{0pt}{\pgfplotmarksize}}%
  \pgfpathlineto{\pgfqpoint{-\pgfplotmarksize}{0pt}}%
  \pgfpathclose\pgfusepathqfillstroke%
}
\pgfdeclareplotmark{thindiamondfilled}{%
  \pgfpathmoveto{\pgfqpoint{0pt}{-\pgfplotmarksize}}%
  \pgfpathlineto{\pgfqpoint{0.6\pgfplotmarksize}{0pt}}%
  \pgfpathlineto{\pgfqpoint{0pt}{\pgfplotmarksize}}%
  \pgfpathlineto{\pgfqpoint{-0.6\pgfplotmarksize}{0pt}}%
  \pgfpathclose\pgfusepathqfillstroke%
}

\pgfplotsset{
  mMWPM/.style={color=black, line width=2.5pt,
    mark=*, mark size=4.5pt,
    mark options={solid, fill=black, draw=black, line width=1.2pt}},
  mBP4M/.style={color=tabpurple, line width=2.5pt,
    mark=plusfilled, mark size=4.5pt,
    mark options={solid, fill=tabpurple, draw=tabpurple, line width=1.2pt}},
  mBP4MF/.style={color=tabcyan, line width=2.5pt,
    dash pattern=on 16.65pt off 7.2pt,
    mark=diamondfilled, mark size=6.364pt,
    mark options={solid, fill=tabcyan, draw=tabcyan, line width=1.2pt}},
  mBP4MM/.style={color=taborange, line width=2.5pt,
   mark =thindiamondfilled, mark size=6.364pt,
    mark options={solid, fill=taborange, draw=taborange, line width=1.2pt}},
  panel/.style={
    width=12.3cm, height=6.7cm,
    scale only axis=false,
    xmin=0.019, xmax=0.16,
    ymin=5e-7, ymax=1,
    xlabel={},
    ylabel={LER},
    label style={font=\fontsize{16}{18}\selectfont},
    title style={font=\fontsize{16}{18}\selectfont},
    tick label style={font=\fontsize{16}{18}\selectfont},
  xtick={0.02,0.06,0.1,0.14},
  scaled x ticks=true,
  xticklabels={$0.02$,$0.06$,$0.1$,
               $0.14$},
    tick align=outside,
    xtick pos=left, ytick pos=left,
    grid=both,
    grid style      ={mplgrid, opacity=0.7, dotted, line width=1pt},
    minor grid style={mplgrid, opacity=0.7, dotted, line width=1pt},
    axis on top=false,
},
}

\begin{axis}[
  ymode=log,
  axis background/.style={fill=white},
  panel,
  name=axtop,
  title={$d = 9$},
  xmode=log,  
]

  \addplot[mMWPM, forget plot] table[x=p, y=ler] {
      p            ler
      0.15         0.244678248
      0.14         0.208986416
      0.13         0.161838485
      0.12         0.116972745
      0.11         0.081083273
      0.1          0.056069526
      0.09         0.034050667
      0.08         0.021060612
      0.07         0.009758859
      0.06         0.004713424
      0.05         0.00185671
      0.04         0.000577
      0.03         0.000131
      0.02         1.2e-05
  };

  \addplot[mBP4M, forget plot] table[x=p, y=ler] {
      p            ler
      0.15         0.329163924
      0.14         0.278009452
      0.13         0.235017626
      0.12         0.176211454
      0.11         0.136798906
      0.1          0.095657165
      0.09         0.067967104
      0.08         0.041644109
      0.07         0.024705999
      0.06         0.012810823
      0.05         0.005896226
      0.04         0.002041746
      0.03         0.000599
      0.02         9.5e-05
  };

  \addplot[mBP4MF, forget plot] table[x=p, y=ler] {
      p            ler
      0.15         0.306748466
      0.14         0.246426811
      0.13         0.201166767
      0.12         0.161186331
      0.11         0.112220851
      0.1          0.081247969
      0.09         0.052562418
      0.08         0.033729088
      0.07         0.017469385
      0.06         0.00901047
      0.05         0.0037134
      0.04         0.001313077
      0.03         0.000313
      0.02         3.7e-05
  };

  \addplot[mBP4MM, forget plot] table[x=p, y=ler] {
      p            ler
      0.15         0.247157687
      0.14         0.202675314
      0.13         0.148
      0.12         0.116849731
      0.11         0.082379109
      0.1          0.055041832
      0.09         0.034575755
      0.08         0.019473438
      0.07         0.01085305
      0.06         0.004902465
      0.05         0.001938067
      0.04         0.000645
      0.03         0.000145
      0.02         1.9e-05
  };

\end{axis}

\begin{axis}[
  ymode=log,
  axis background/.style={fill=white},
  panel,
  name=axbot,
  at={(axtop.below south west)}, anchor=above north west, yshift=0cm,
  title={$d = 11$},
  xlabel={physical error rate},
  xmode=log,  
  legend style={
    at={(0.9754,0.0414)}, anchor=south east,
    font=\fontsize{14}{16}\selectfont,
    fill=white, fill opacity=0.9, text opacity=1,
    draw=black!25, rounded corners=2pt,
    row sep=1pt,
  },
  legend cell align=left,
  legend image code/.code={%
    \draw[mark repeat=2, mark phase=2, #1]
      plot coordinates {(0cm,0cm) (0.75cm,0cm) (1.5cm,0cm)};%
  },
]

  \addplot[mMWPM, forget plot] table[x=p, y=ler] {
      p            ler
      0.15         0.248
      0.14         0.190694127
      0.13         0.139043382
      0.12         0.101657009
      0.11         0.065616798
      0.1          0.041430169
      0.09         0.022698384
      0.08         0.011190564
      0.07         0.005175849
      0.06         0.002056788
      0.05         0.00069
      0.04         0.000146
      0.03         1.9e-05
      0.02         1e-06
  };

  \addplot[mBP4M, forget plot] table[x=p, y=ler] {
      p            ler
      0.15         0.366166239
      0.14         0.311623559
      0.13         0.253871541
      0.12         0.193423598
      0.11         0.143988481
      0.1          0.102679947
      0.09         0.06912761
      0.08         0.043084877
      0.07         0.022500225
      0.06         0.01225295
      0.05         0.00529361
      0.04         0.001767481
      0.03         0.000471
      0.02         6.9e-05
  };

  \addplot[mBP4MF, forget plot] table[x=p, y=ler] {
      p            ler
      0.15         0.317965024
      0.14         0.264061262
      0.13         0.199720391
      0.12         0.153988297
      0.11         0.107770234
      0.1          0.071275837
      0.09         0.044155959
      0.08         0.026174584
      0.07         0.012719895
      0.06         0.005669031
      0.05         0.002079711
      0.04         0.000642
      0.03         0.000113
      0.02         8e-06
  };

  \addplot[mBP4MM, forget plot] table[x=p, y=ler] {
      p            ler
      0.15         0.255689082
      0.14         0.196540881
      0.13         0.143
      0.12         0.09544717
      0.11         0.064424688
      0.1          0.040559724
      0.09         0.023073906
      0.08         0.011709465
      0.07         0.005261386
      0.06         0.002104723
      0.05         0.000638
      0.04         0.000159
      0.03         2.9e-05
      0.02         2e-06
  };

  \addlegendimage{mMWPM}   \addlegendentry{MWPM}
  \addlegendimage{mBP4M}   \addlegendentry{BP4M}
  \addlegendimage{mBP4MF}  \addlegendentry{BP4MF}
  \addlegendimage{mBP4MM}  \addlegendentry{BP4M+M}

\end{axis}
\end{tikzpicture}}
\captionof{figure}{Code capacity performance of our algorithms (BP4M, BP4MF,
BP4M+M) with 25 iterations compared to MWPM for distances $d=9,11$.}
\label{fig: performance comparison}
\end{minipage}%
\vspace{1 cm}
\hfill%
\begin{minipage}[t]{\columnwidth}
\centering
\resizebox{\linewidth}{!}{
\begin{tikzpicture}

%
\newcommand{\cpms}{4.5pt}
\definecolor{mplpink}{HTML}{008080}

\pgfplotsset{
  cpcurve/.style={
    color=#1,
    line width=3.5pt,
    mark=*,
    mark size=\cpms,
    mark options={solid, fill=#1, draw=#1, line width=1.2pt},
    error bars/y dir=both,
    error bars/y explicit,
    error bars/error bar style={color=#1, line width=2.5pt},
    error bars/error mark=-,
    error bars/error mark options={color=#1, mark size=3pt, line width=2.5pt},
  },
  cpkey/.style={
    color=#1, line width=3.5pt, mark=*, mark size=\cpms,
    mark options={solid, fill=#1, draw=#1, line width=1.2pt},
  },
}

\begin{axis}[
  xmode=log, ymode=log,
  axis background/.style={fill=white},
  width=22.86cm, height=15.24cm,          
  scale only axis=false,
  xmin=0.0018286156535236553, xmax=0.01312468257271731,
  ymin=2.9341616804130624e-05, ymax=0.5462227573011053,
  xlabel={physical error rate },
  ylabel={},
  label style={font=\fontsize{25}{30}\selectfont},
  tick label style={font=\fontsize{20}{24}\selectfont},
  xtick={0.002,0.003,0.004,0.006,0.01},
  xticklabels={$2\times10^{-3}$,$3\times10^{-3}$,$4\times10^{-3}$,
               $6\times10^{-3}$,$10^{-2}$},
  minor xtick={0.005,0.007,0.008,0.009},
  ytick={1e-4,1e-3,1e-2,1e-1},
  minor ytick={3e-05,4e-05,5e-05,6e-05,7e-05,8e-05,9e-05,0.0002,0.0003,
              0.0004,0.0005,0.0006,0.0007,0.0008,0.0009,0.002,0.003,
              0.004,0.005,0.006,0.007,0.008,0.009,0.02,0.03,0.04,0.05,
              0.06,0.07,0.08,0.09,0.2,0.3,0.4,0.5},
  tick align=outside,
  xtick pos=left, ytick pos=left,
  grid=both,
  grid style      ={mplgrid, dotted, line width=1pt},
  minor grid style={mplgrid, dotted, line width=1pt},
  axis on top=false,
  legend style={
    at={(0.98,0.02)}, anchor=south east,
    font=\fontsize{18}{22}\selectfont,
    fill=white, fill opacity=0.9, text opacity=1,
    draw=black!25, rounded corners=2pt,
    row sep=1pt, column sep=8pt,
  },
  legend columns=2,
  legend cell align=left,
  legend image code/.code={%
    \draw[mark repeat=2, mark phase=2, #1]
      plot coordinates {(0cm,0cm) (0.55cm,0cm) (1.1cm,0cm)};%
  },
]

  \addplot[forget plot, mplbrown, mpldotted, line width=3.5pt]
    coordinates {(0.0057,2.9341616804130624e-05) (0.0057,0.5462227573011053)};
\addplot[forget plot, black,    mpldotted, line width=3.5pt]
    coordinates {(0.0072,2.9341616804130624e-05) (0.0072,0.5462227573011053)};

  \addplot[cpcurve=mplgreen, forget plot]
    table[x=p, y=ler, y error plus=eplus, y error minus=eminus] {
      p        ler              eplus            eminus
      0.002    0.001244406098   8.882891161e-05  8.443472448e-05
      0.003    0.003512573242   0.0002388491416  0.0002275901356
      0.004    0.007591610863   0.0004826588985  0.0004614005597
      0.005    0.0143761268     0.0008392853821  0.0008054388968
      0.006    0.02298583984    0.00120134998    0.00115816121
      0.007    0.03428649902    0.001627238381   0.001574560786
      0.008    0.04565429688    0.001991899881   0.001933192371
      0.009    0.06372070312    0.002512494942   0.002446757044
      0.01     0.08312988281    0.002832714984   0.002769941794
      0.011    0.1020100911     0.003100009669   0.003040105369
      0.012    0.1226806641     0.003679587735   0.003611433345
    };

  \addplot[cpcurve=mplpurple, forget plot]
    table[x=p, y=ler, y error plus=eplus, y error minus=eminus] {
      p        ler              eplus            eminus
      0.002    0.0002026171654  1.484051903e-05  1.40875359e-05
      0.003    0.001440942768   0.0001049836263  9.968978952e-05
      0.004    0.004287791913   0.0003092415508  0.0002938393008
      0.005    0.0110436793     0.0007765796894  0.0007390657063
      0.006    0.02353232598    0.001590759052   0.001517665447
      0.007    0.04275334112    0.002654030691   0.002543864854
      0.008    0.06930063579    0.003931486194   0.003786281974
      0.009    0.100198167      0.005418998931   0.005235220923
      0.01     0.1375264911     0.006855159997   0.006651576779
      0.011    0.1839237057     0.008154719032   0.007955093661
      0.012    0.2336512262     0.009499979521   0.009307762602
    };

  \addplot[cpcurve=mplpink, forget plot]
    table[x=p, y=ler, y error plus=eplus, y error minus=eminus] {
      p        ler              eplus            eminus
      0.002    5.079051741e-05  5.289260907e-06  4.916185036e-06
      0.003    0.0006424113553  4.704175605e-05  4.465647004e-05
      0.004    0.003153554574   0.0002300201809  0.0002184280001
      0.005    0.01115517016    0.0008023545656  0.0007627606575
      0.006    0.02547632794    0.00178691992    0.001702054357
      0.007    0.05273287144    0.003579128403   0.003419157713
      0.008    0.09508640599    0.006026941695   0.005787292476
      0.009    0.1465357968     0.00872641712    0.008423101106
      0.01     0.2149345651     0.01166249064    0.01133636801
      0.011    0.269319595      0.01349058564    0.01318610737
      0.012    0.3338232207     0.01554554407    0.01528628732
    };

  
  \addplot[forget plot, mplbrown, mpldashdot, line width=3.5pt,
           domain=0.0018286156535236553:0.01312468257271731, samples=2] {x};

  \node[anchor=north west,
        draw=black!25, fill=white, fill opacity=0.9, text opacity=1,
        rounded corners=2pt, inner sep=6pt,
        font=\fontsize{18}{22}\selectfont]
    at (rel axis cs:0.02,0.98) {Rotated Surface memory-Z};

  \node[anchor=south east, mplbrown,
        font=\bfseries\fontsize{20}{24}\selectfont]
    at (axis cs:0.00560, 3.43e-4) {B-BP4MF};
  \node[anchor=south east, mplbrown,
        font=\bfseries\fontsize{20}{24}\selectfont]
    at (axis cs:0.00560, 6.82e-4) {0.0057};
  \node[anchor=south west, black,
        font=\bfseries\fontsize{20}{24}\selectfont]
    at (axis cs:0.00740, 3.43e-4) {MWPM};
  \node[anchor=south west, black,
        font=\bfseries\fontsize{20}{24}\selectfont]
    at (axis cs:0.00740, 6.82e-4) {0.0072};


\end{axis}
\end{tikzpicture}}
\captionof{figure}{B-BP4MF threshold result for circuit-level noise rotated
memory-Z surface code with 75+75 iterations.}
\label{fig: threshold bp4+bp4mf}
\end{minipage}

\vspace{\dbltextfloatsep}

\end{document}